# One-Dimensional Structural Properties of Proteins in the Coarse-Grained CABS Model

**Sebastian Kmiecik and Andrzej Kolinski***

Faculty of Chemistry, University of Warsaw, Pasteura 1, 02-093 Warszawa, Poland

*email: kolinski@chem.uw.edu.pl

**Abstract**

Despite the significant increase in computational power, molecular modeling of protein structure using classical all-atom approaches remains inefficient, at least for most of the protein targets in the focus of biomedical research. Perhaps the most successful strategy to overcome the inefficiency problem is multiscale modeling to merge all-atom and coarse-grained models. This chapter describes a well-established CABS coarse-grained protein model. The CABS (C-Alpha, C-Beta and Side chains) model assumes a 2-4 united-atom representation of amino acids, knowledge-based force field (derived from the statistical regularities seen in known protein sequences and structures) and efficient Monte Carlo sampling schemes (MC dynamics, MC replica-exchange, and combinations). A particular emphasis is given to the unique design of the CABS force-field, which is largely defined using one-dimensional structural properties of proteins, including protein secondary structure. This chapter also presents CABS-based modeling methods, including multiscale tools for *de novo* structure prediction, modeling of protein dynamics and prediction of protein-peptide complexes. CABS-based tools are freely available at http://biocomp.chem.uw.edu.pl/tools

# 1 Introduction

In the last two or three decades, we are witnessing incredible progress in experimental and theoretical molecular biology. Thanks to intensive experimental studies, especially genome projects, huge amounts of sequence data (primary structures of proteins, nucleic acids and other biomacromolecules) are now available. The combination of new experimental techniques and theoretical tools for their interpretation also provides structural (three dimensional) data for many biological macromolecules. Nevertheless, experimentally determined structures remain unknown for an increasing fraction of known protein sequences (but also other biomacromolecules). The explanation of this growing gap is simple: sequencing is now easier, faster and less expensive than structure determination.

Deeper understanding of the molecular basis of life processes requires not only determination of structures of single biomacromolecules but also realistic pictures of their interaction with other biomacromolecules, mechanisms of assembly processes and structural and dynamic properties of resulting complexes. Taking into consideration that we know experimental structures of only a fraction of monomeric proteins, and that the estimated number of possible protein dimers (oligomers) is an order of magnitude larger than the number of monomers, it becomes obvious that structural biology needs strong support from theoretical studies. Efficient methods for structure prediction, and modeling of dynamics and interaction are necessary. Many important problems of molecular biology can be studied using classical all-atom molecular dynamics (MD) methods. For very small and fast folding proteins it is now possible to simulate the entire folding process using superfast dedicated computers [1]. For larger systems it is still beyond capability of the available computing technology, and the time gap is huge. This is the main reason for the development of new molecular modeling tools that can handle large systems. Simplified coarse-grained, and thereby computationally very fast, models can be used for simulations of large biomacromolecules and/or for the modeling of long time processes [2, 3]. Useful coarse-grained models need to be of sufficient resolution, enabling reasonable connection with atomistic pictures [3, 4]. The high importance of such methods has been recognized a long time ago [5], resulting in the plethora of new molecular modeling tools. Recently, "*the development of multiscale models for complex chemical systems*", a pioneering work of Karplus, Levit and Warshel, was awarded the Nobel Prize in Chemistry for 2013.

Several very efficient coarse-grained protein models have been developed. Some of them, such as Rosetta [6, 7] or I-Tasser [8, 9], are targeted onto structure prediction, while others, such as CABS [10, 11] or UNRES [12, 13], are more universal, enabling not only structure modeling but also realistic simulations of the dynamic properties of protein systems.

The methods presented in this chapter are based on the CABS (C-Alpha, C-Beta and Side chains) discrete representation of protein chains. Two quite fundamental features make CABS qualitatively different from other coarse grained models. The first one is that the coordinates of the model chains are restricted to discreet positions in a simple three dimensional lattice. Lattice spacing is small enough to ensure good resolution of chain representation, and large enough to make possible predefinition (and storage as integer numbers in large data tables) of all possible local conformations. This way, due to the simple computation of local moves and related energy changes, the Monte Carlo dynamics simulations are much faster than it would be possible for otherwise equivalent continuous models. The second unique feature of CABS is its interaction scheme. The force field consists of knowledge-based statistical potentials derived from the regularities observed in the known protein structures. All interactions, especially those between side chains, are treated as context-dependent. This way complicated multi-body effects are encoded in pairwise potentials. The potentials describing the energies of side chain-side chain interactions depend on the secondary structure of the interacting fragments, their mutual orientations and on the distance (short distance contacts only are treated in the explicit fashion) between side chain centers. Such a context-dependent model of pairwise interactions, especially its dependence on secondary structure, encodes the averaged effects of many physical interactions, and all these interactions, including complex interactions with the solvent, are treated in an implicit fashion.

The coarse-graining level of the CABS model enables fast and realistic reconstruction of atom level structure representation, enabling efficient multiscale modeling of protein systems [4]. The CABS model has proven to be a good tool for the computational prediction of three-dimensional protein structures, including *de novo* and comparative modeling, studies of protein dynamics and folding pathways, and flexible docking.

This chapter is organized as follows. In the **Materials** section we describe the CABS protein structure representation, its force field and the sampling method. Special attention is given to the context-dependent force field of the model, which is strongly dependent on the one-dimensional properties of protein chains, especially their secondary structure assignments.

In the **Methods** section, we list and briefly describe various protein modeling methods based on the CABS model with the emphasis on those utilizing sequence and secondary structure data only. These methods include publicly available modeling tools: CABS-fold: server for protein structure, including *de novo*, modeling and comparative modeling using one or more structural analogs [14]; CABS-dock: server for the flexible docking of peptides to proteins using no knowledge about the binding site [15, 16]; and pyCABS: software package for the simulation and analysis of long-term protein dynamics of globular proteins [17]. In the **Case studies** section, we present example performance of CABS-fold, CABS-dock and pyCABS, together with short descriptions of their input requirements and options.

Finally, the **Notes** section provides several specific comments about the modeling results obtained using CABS-based methods, their further utilization, interpretation, or alternative modeling techniques that may enhance modeling accuracy.

## 2. Materials

### 2.1 CABS model: coarse grained representation of protein structure

The CABS model is a universal tool for the modeling of protein structure dynamics and protein molecular docking. The main chain of protein structure is represented by a chain of C$\alpha$-atoms and pseudo atoms representing the center of virtual C$\alpha$-C$\alpha$ bonds (see Fig. 1A and B). The latter one is needed for the simplified definition of hydrogen bonds. Side chains are represented by C$\beta$ atoms and pseudo-atoms representing the centers of the remaining portions (where applicable) of the amino acid side chains. The CABS C$\alpha$ trace is placed onto a lattice network with 0.61 Å spacing. This lattice representation significantly speeds up the Monte Carlo sampling scheme when compared with continuous models of similar resolution. The lattice spacing of 0.61 Å enables a large set of allowed orientations of C$\alpha$-C$\alpha$ virtual bonds (when a slight fluctuation of their length is allowed) and thereby eliminates any noticeable orientation biases that are present in simple lattice models. The average accuracy of the C$\alpha$-trace representation is about 0.35 Å, and slightly depends on the secondary structure patterns of the proteins studied (see Fig. 1C).

### 2.2 Force field of the CABS model with secondary structure context-dependent statistical potentials

The force field of CABS is constructed from knowledge-based statistical potentials, derived from the structural regularities (and their relation to the amino acid sequences) seen in protein structures collected in databases. A large representative set has been used for the derivation of all potentials. The weight of various potentials is properly tuned by optimizing the total energy of folded structure and other properties of the model, for instance secondary structure content at folded and unfolded structures of the proteins being modeled. The details of the force field of CABS models and the motivations for specific choices of their potentials have been described previously [10]. Here we outline the general ideas behind this force field, focusing on the crucial role of secondary structure assignments for the model and its force field.

Protein chain geometry in the CABS model is fully encoded by its C$\alpha$-trace, where positions of all C$\alpha$ atoms are restricted to the points of the underlying cubic lattice grid. The planar angles between two subsequent C$\alpha$-C$\alpha$ pseudo-bonds are restricted to values seen in protein structures. Sequence-independent and sequence-dependent potentials enforce distribution of this angle typical for the distribution seen in globular proteins. The angles of rotation of three consecutive C$\alpha$-C$\alpha$ pseudo-bonds are similarly treated. This way, for instance, left-handed helix-like conformations are treated as unlike. The sequence dependence of the angular potentials is not straightforward, and it does not come from a specific identity of three or four residue fragments, but from the predicted secondary structure which depends on much longer protein fragments. This way complex multibody interactions are encoded in this simple potential. Positions of C$\beta$ carbons (not restricted to the lattice) are defined by the positions of three consecutive C$\alpha$ atoms for the C$\beta$ bound to the central C$\alpha$. These positions depend on the planar angles between the C$\alpha$-C$\alpha$ pseudo-bonds. C$\alpha$ and C$\alpha$ united atoms are treated as rigid bodies. Virtual united atoms, placed at the center of atom-C$\alpha$ pseudo-bonds, define the positions of the main chain hydrogen bonds in the form of attractive, orientation-dependent contact potentials of the same strength for all residues belonging to the same pre-defined secondary structure assignment. Other hydrogen bonds have the same geometry, but are considered weaker, with a smaller weight factor. The excluded volume spheres of united C$\alpha$, C$\beta$ atoms and hydrogen bonds forming pseudo-atoms are slightly smaller in the CABS model than the distance of a corresponding strong repulsive interaction in real proteins. This is necessary to enable non-perfect dense packing in the native-like structure of the modeled proteins. Positions of centers of the remaining portion of amino acid side chains (where applicable) are also taken from tables defined for specific amino-acids and local angles of the C$\alpha$-trace. Interactions between the centers of amino acid side chains are most important for the performance of the CABS model. Side chains are treated as soft excluded volume bodies at short distances, and interacting through contact potential at a longer distance. The width of contact distance is about 2 Å. The soft excluded volume of the side chains and the width of the contact range cover the potential problems with non-accurate representation of side chain conformations, especially for larger amino acids.

Side chain pairwise contact potentials are crucial for the performance of the CABS model. These statistical potentials are context dependent, and the strength of pairwise interactions depends on the mutual orientation of the interacting side chains and on the geometry of the nearest fragments of the main chain backbone. Here we discuss and present this potential for single domain globular proteins. It is important to note that the reference state in the derivation of CABS statistical potentials is a compact state of protein chains with a random sequence of the same composition as the protein of a given composition. Similar context-dependent potentials can be derived for interactions between globular proteins, trans-membrane proteins, etc. It means that the CABS force field is not easily "transferable", it is rather "expandable" for an increasing range of modeled systems. We do not consider this a strong disadvantage of "knowledge-based" statistical potentials. "Transferability" of "physics-based" force fields for reduced models is also not trivial [18].

The idea of the context-dependent classification of side chain contacts is illustrated in Figure 2 and numerical data are presented in Tables 2 to 10. The mutual orientation of the contacting side chains is divided into three ranges: near-antiparallel, intermediate and near parallel (Fig. 2a). The local geometry of the main chain of a contacting residue is classified in the CABS force field into two classes: compact and expanded. This way the secondary structure prediction (or assignment) defines the specific energy of side chain interactions. For example: an antiparallel contact of two residues with a compact geometry of the corresponding elements of the main chain backbone usually means a helix-helix contact, while a parallel contact of side chains from two expanded elements of the main chain usually comes from two adjoined beta strands. The context-dependent contact potentials of the CABS force field differ qualitatively from the other potentials (probably all of them) used in protein modeling. In the majority of these potentials the contact energy for two oppositely charged amino-acids will suggest weak attractive interactions of their side chains. In the CABS force field (for single-domain globular proteins) interactions of such pairs of residues are treated as strongly attractive for a parallel contact and strongly repulsive for an antiparallel contact of the side chains. Since the solvent in the CABS model force field is treated in a strictly implicit fashion (which is also the case for the majority of other statistical potentials) such orientation-dependent strength of interactions is not surprising. Charged residues are usually located on the surface of a protein globule, where they cannot form antiparallel contacts. If they are located (which is rather rare) more in the center of a globule, it is most likely that it is a binding site, where charged residues are on the surface of the binding site. Also in this case the parallel contact is more probable.

As discussed above, the predicted (or assigned) secondary structure in a three-state version (helix, beta, other) is crucial for CABS force field statistical potentials. This unique feature of the CABS coarse-grained modeling approach is a strength of the model, with very few drawbacks. The model has proved to be very efficient in *de novo* protein structure assembly simulations, comparative modeling support, modeling of protein dynamics and interactions with other biomolecules. In the last case the force field needs to be properly expanded, including for example contact potentials between side chains from two protein (peptide) chains. Due to the qualitative difference between CABS side chain contact potentials and other statistical potentials, we decided to attach its numerical data presented in twelve tables (numbered from 2 to 10). Two-digit accuracy is sufficient for most applications of this potential.

The contact potentials data (Tables 2-10) are potentially very useful not only in coarse-grained modeling (with model resolution similar to that assumed in the CABS model) but also as a source for definition/sorting of many other one-dimensional, two-dimensional and three-dimensional protein features. For instance it is possible to use the numerical data of this potential for the classification of burial patterns of protein sequences. The potential can also be used in efficient threading algorithms, and in other structural bioinformatics methods. Additional comments on the meaning of the tables and accessibility in software packages are provided in Note 1.

**2.3 Sampling schemes**

The Monte Carlo sampling scheme of CABS is a series of local, randomly selected, small conformational transitions onto the underlying lattice. The set of local changes of model chain coordinates includes single C$\alpha$ moves (see Fig. 1B), moves of two C$\alpha$ fragments, and rarely attempted small distance moves of longer fragments of the model chains. Chain ends are treated separately. Due to lattice discretization of the C$\alpha$ coordinates (800 of allowed orientations of C$\alpha$-C$\alpha$ pseudo-bond vectors) the possible local moves could be stored in large data tables and thereby local moves do not require any costly computations of trigonometric functions. Local moves require just simple random sorting of predefined sequences of integer numbers. This way the discrete (restricted to a high coordination lattice) representation of chain conformations makes the CABS model computationally much faster in comparison to otherwise equivalent continuous coarse-grained models. The geometry of the main chain defines the positions of the side chain united atoms (not restricted to the lattice). A library of these positions is pre-defined by sorting and averaging PDB structures for all possible amino acid sequences of the central and two neighboring elements of the C$\alpha$–trace. All random moves are accepted according to the Metropolis criteria. Since the randomly selected moves mimic fast local conformational fluctuations of the modeled protein chains, their long series provides a realistic picture of the long time dynamics of modeled systems. CABS-based modeling schemes can use simple MC dynamics simulations at a given temperature, MC simulated annealing, and various versions of Replica Exchange (REMC) simulations. The CABS model (MC dynamics or REMC) could be easily combined with all-atom molecular dynamics. Several simple algorithms, classical and specifically targeted onto CABS representation, can be used for the fast and realistic reconstruction of atomistic representation, suitable for classical MD simulations (see Note 2). This way CABS can be used as a very efficient engine in multiscale protein modeling schemes. The basic structure of multiscale modeling procedures with CABS is illustrated in Fig. 3. Some helpful tools for the analysis of derived models and CABS trajectories are presented in Note 3 and 4.

## 3 Methods

In the last several years, CABS coarse-grained protein models have become a key component in various multiscale modeling methods. Those methods generally follow a similar pipeline merging CABS simulations (usually the first modeling step) and all-atom modeling (final modeling steps), as presented in Figure 3.

The CABS-based modeling methods have three application areas:

(1) protein structure prediction: homology modeling [14, 19-21], *ab initio* prediction of small proteins [14] or protein loops/fragments [22-24] (in [23] also in combination with the classical Modeller tool [25]), modeling based on sparse experimental data [26]

(2) prediction of protein complexes: protein-peptide [15, 16, 27] and protein-protein [28, 29]

(3) efficient simulation of protein dynamics: protein folding mechanisms [4, 11, 30-34] and flexibility of globular proteins [35-38]

In all these applications, the CABS model serves as a highly efficient simulation engine that allows CABS-based methods to be much cheaper in terms of CPU time (in comparison to classical modeling tools [35]), or to achieve sampling efficiency that exceeds other existing approaches. For example, the CABS-dock method for the molecular docking of peptides to proteins enables docking fully flexible peptides to flexible receptors without prior knowledge of the binding site [15, 16]. In practice, CABS-dock performs simulation of coupled folding and binding during which peptides have a possibility to explore the entire surface of a protein receptor. Presently, there are no other simulation methods enabling exploration of such a large conformational space in a reasonable time. In contrast to CABS-dock, other state-of-the-art protein-peptide docking methods are restricted to a specified binding site, or to very short peptides (2-4 amino acids, while CABS-dock has been successfully tested on a large set of peptides with 5–15 amino acids [15, 16]).

In Table 1, we list the CABS-based methods that enable protein structure modeling based on one-dimensional data only (sequence and secondary structure), together with their accessibility, references, benchmark information and performance summary. For selected methods (CABS-fold [14], CABS-dock [15, 16] and pyCABS [17]), example case studies are presented in the next section.

**Table 1**. Performance of the CABS-based modeling methods in *ab initio* prediction tasks (utilizing one dimensional data only: protein sequence and secondary structure).

| *Method and availability* | *Benchmark set* | *Performance summary* |
|---|---|---|
| **Prediction of protein structure or protein fragments** | | |
| CABS-fold server for the *ab initio* and consensus-based prediction of protein structure [14]. Available as a web server at: http://biocomp.chem.uw.edu.pl/CABSfold/ | Methodology validated during CASP competitions as one of the leading approaches [19-21], applied to the *ab initio* modeling of large protein fragments or entire proteins (with or without 3D restraints). | Small proteins (up to 100 residues long) or peptides can be predicted with high accuracy (up to 2 Å) or medium accuracy (up to 5 Å).<br><br>The CABS-fold server can also be used to predict protein loops (see the performance below). |
| Method(s) for predicting protein loops in globular proteins [23]. | From 186 experimental protein structures, covering all the structural classes of proteins, internal loops of various length (from 4 to 25 residues) have been removed and treated as unknown. | Performance was compared with two classical modeling tools: Modeller [25] and Rosetta [6]. Modeller performance was usually better for short loops, while CABS and Rosetta were more effective |
| | | for longer loops (resolution of such models was usually on the level of 2-6 Å). |
| Prediction method for protein loops in GPCR membrane receptors [24]. | From 13 experimental GPCR receptor structures, extracellular second loops (between 13 and 34 residues) have been removed and treated as unknown. The benchmark set is available at: http://biocomp.chem.uw.edu.pl/GPCR-loop-modeling/ | Resolution of the best models obtained (among many others) was on the level of 2-6 Å, while the best scored models were on the level of 2-8 Å. Performance was comparable to that of other state-of-the-art methods [24]. |
| Method for protein fragment reconstruction [22]. | From 20 protein structures of various structural classes, protein fragments (from 10 to 29 residues) have been removed and treated as unknown. | Resolution of the resulting models was on the level of 1.5 and 6 Å. Performance was compared with SICHO [39], Refiner [22], Swiss-model [40] and Modeller [25] methods. CABS, SICHO and Refiner performance was usually better than for Swiss-model and Modeller. |
| **Protein-peptide molecular docking and binding site prediction (using no knowledge about the peptide structure)** | | |
| CABS-dock method for molecular docking with no knowledge of the binding site [15, 16]. Available as a web server at: http://biocomp.chem.uw.edu.pl/CABSdock/ | Benchmark set of non-redundant (<70% sequence identity with respect to the receptor protein) protein-peptide interactions (108 bound and 68 unbound receptors) with peptides of 5–15 amino acids [41]. The benchmark set is available at: http://biocomp.chem.uw.edu.pl/CABSdock/benchmark | For over 80% of bound and unbound cases high or medium accuracy models were obtained (high accuracy: peptide-RMSD<3 Å; medium accuracy: 3 Å ≤ peptide-RMSD ≤ 5.5 Å; where peptide-RMSD is the RMSD to the experimental peptide structure after superimposition of receptor molecules. |
| *Ab initio* protocol for studying the folding and binding mechanism of intrinsically disordered peptides [27]. | pKID-KIX protein complex (pKID is a 28 residue disordered peptide which folds upon binding to the KIX domain). | An ensemble of transient encounter complexes obtained in the simulations was in good agreement with experimental results. |
| **Prediction of protein folding mechanisms** | | |
| pyCABS protocols for efficient simulations of long-time protein dynamics [17]. Software package available at: http://biocomp.chem.uw.edu.pl/pycabs/ | Tested in protein folding studies of small (up to 100 residues long) globular proteins [4, 11, 30-33]. | The views of denatured ensembles of protein structures obtained in the simulations were in good agreement with the experimental measurements of protein folding [4, 11, 30-33]. |
| Multiscale protocol merging efficient simulations with CABS and replica exchange all-atom MD [34]. | β-hairpin from the B1 domain of protein G (PDB code: 2GB1, residues 41 to 56). | Combination of CABS and all-atom MD simulations significantly accelerates system convergence (several times in comparison with all-atom MD starting from the extended chain conformation) |

Apart from the methods listed in Table 1, the CABS model has also been used in web server tools: CABS-flex server for the prediction of protein structure fluctuations [36, 37] and Aggrescan3D server for the prediction of protein aggregation properties and rational design [38] (Aggrescan3D uses the CABS-flex method for modeling the influence of conformational flexibility on aggregation properties). The major advantage of the CABS-flex method is its efficiency. It allows us to achieve similar results as with classical all-atom MD, but several thousand times faster [35].

**4 Case Studies**

### 4.1 Protein structure prediction using the CABS-fold server

The CABS-fold server for protein structure prediction operates in two modeling modes: consensus modeling (based on structural templates) and *de novo* modeling (based only on sequence) [14]. In both modes, the secondary structure is an optional input (see Note 5): if the secondary structure is not provided, it is automatically predicted using the Psi-Pred method [42]. It is also possible to add distance restraints into the modeling process and to modify CABS simulation settings. These additional options can be accessed from the "Advanced options" input panel (see Figure 4 presenting example CABS-fold screenshots).

CABS-fold performance and the benchmark summary are presented in Table 1. In Figure 5, we present an example modeling result using the *de novo* modeling mode and the sequence of a small protein domain, yeast copper transporter CCC2A (72 residues). Protein sequence and secondary structure inputs are also provided in the figure. The CCC2A protein structure has been solved experimentally and has a beta-alpha-beta-beta-alpha-beta ferrodoxin-like fold (PDB ID: 1fvq). Figure 5 shows a comparison of the experimental and CABS-fold predicted model with the same fold which differ in details of secondary structure packing. It is worth to mention that obtaining such a modeling result based on protein sequence only is not trivial and possible (in a reasonable computational time) only using a few coarse-grained based methods.

### 4.2 Protein-peptide docking using the CABS-dock server

The CABS-dock server for modeling protein-peptide interactions [15, 16] enables efficient docking search of a peptide over the entire protein receptor structure. During CABS-dock docking, the peptide is simulated as fully flexible, while the protein receptor structure is also flexible but only to a small extent. As an input, the CABS-dock method uses information about the peptide sequence and structure of a protein receptor. The peptide secondary structure is an optional input (see Note 5; if not provided, the method uses the PsiPred tool [42] for secondary structure prediction). Other optional inputs include the possibility to assign high flexibility for selected receptor fragments, and to exclude selected receptor fragments from docking search

(these are accessible from the optional input panel, see the CABS-dock screenshots in Figure 6).

CABS-fold performance and the benchmark summary are presented in Table 1. In Figure 7, we present an example modeling result obtained using the optional CABS-dock feature that allows for the significant flexibility of a selected receptor fragment. In the presented modeling case, assigning significant flexibility to the flexible loop (which partially blocks the binding site in the unbound input form) was crucial for obtaining a high resolution complex model.

### 4.3 Protein dynamics using the pyCABS package

The pyCABS software package [17] is dedicated to performing long-time simulations of small globular proteins using the CABS model. The possible applications include *de novo* folding from a random structure (folding mechanisms), near-native dynamics, unfolding processes, and long-time dynamics of unfolded structures. The package requires the protein sequence and its secondary structure (predicted or experimentally assigned, see Note 5) and starting structure(s): depending of the modeling goal, it can be a random structure, or a selected (e.g. native) structure.

pyCABS performance and the benchmark summary are presented in Table 1. In Figure 8, we present an example modeling result from the simulation of folding of barnase globular protein. The simulation was performed in the *de novo* manner, i.e. using a random starting structure. The resulting picture of the folding mechanism matches well the experimental data and has been described in detail in [11] (the technical details for carrying out such a simulation using pyCABS are provided in [17]).

## 5 Notes

1. Context-dependent contact potentials in CABS software packages.

   Tables 2-10 are an integral part of the CABS (and pyCABS [17]) software package (stored in the "QUASI3S" text file). Each of these tables is labeled by a three letter code (like: PEE, PCC, PCE) whose meaning is explained in Fig. 2. For example, the PCE type of interactions occurs between amino acid chains forming a parallel contact (P), where the first contacting side chain (given in columns) is attached to a compact (C, most likely a helix) type of conformation and the second contacting side chain (given in rows) is attached to expanded (E, most likely beta-strand) conformation.

2. Tools for reconstruction from $C\alpha$ atom positions to all-atom representation and further optimization.

The basic output of the CABS model is a trajectory in $C\alpha$ representation. The CABS coarse-grained trajectories, or selected trajectory models, can be reconstructed to all-atom representation. The major output of the CABS-based multiscale methods (like CABS-fold or CABS-dock servers) is a set of a few models in all-atom representation (automatically selected and reconstructed). These methods also provide hundreds (CABS-fold) or thousands (CABS-dock) of predicted models in $C\alpha$ trajectories that may be useful in a more thorough analysis of the prediction results and reconstructed to all-atom resolution by the user.

There are many strategies for the reconstruction from the $C\alpha$ to all-atom format; however, the method chosen should be insensitive to small local distortions of the C-alpha distances present in CABS-generated models. Based on our experience, we can recommend the following reconstruction protocols:

- ModRefiner package [43] for combined reconstruction and optimization (handles only monomeric protein chains, employed in the CABS-fold [14] server).
- Modeller package [44] for combined reconstruction and optimization (employed in the CABS-dock server, details of the Modeller protocol are provided in [16] and the CABS-dock online tutorial http://biocomp.chem.uw.edu.pl/CABSdock/tutorial).
- Claessens et al. [45] or BBQ [46] approach for protein backbone reconstruction followed by the second rebuilding step (side chain reconstruction) using the SCWRL program [47].

The last two-step protocols require a third additional optimization step, which is more demanding when BBQ is used for backbone reconstruction [48]. We tested the performance of such reconstruction and fast optimization protocols in protein structure prediction [48] and protein dynamics [30] exercises. Optimization strategies have also been reviewed in [49].

3. Models scoring

Reconstructed and optimized all-atom models can be assessed using specially designed scoring methods. An accurate scoring function that can discriminate near-native models, or docking poses, from a large set of alternative solutions is an important component of structure prediction methodologies [50-52].

4. Trajectory analysis

CABS modeling trajectories can be additionally analyzed using external tools for the structural clustering and comparison of protein models, e.g. the ClusCo package [53] or hierarchical clustering within the Bioshell package [54]. Convenient analysis of protein models usually requires superimposition of the compared models, or entire trajectories; a useful tool for that is the Theseus package [55].

5. Secondary structure input

The accuracy of *de novo* structure prediction by CABS-fold or CABS-dock servers depends on the accuracy of the secondary structure input. Small errors in the predicted secondary structure do not impose any serious problems, but it is (on average) safer to use underestimated ranges of regular (helices and beta strands) secondary structure fragments than overestimated ranges (for instance prediction of a single long helix for the fragment that forms two differently oriented helices). Qualitative errors of secondary structure predictions, where helical fragments are predicted as beta strands (or vice versa), are dangerous for modeling results. Fortunately, this kind of errors is rare for good bioinformatics tools for secondary structure prediction and could be eliminated by rejecting more problematic predictions.

**Acknowledgments**

Funding for this work was supported by the National Science Center grant [MAESTRO 2014/14/A/ST6/0008] and by the Foundation for Polish Science TEAM project (TEAM/2011-7/6) co-financed by the EU European Regional Development Fund operated within the Innovative Economy Operational Program.

**Figures**

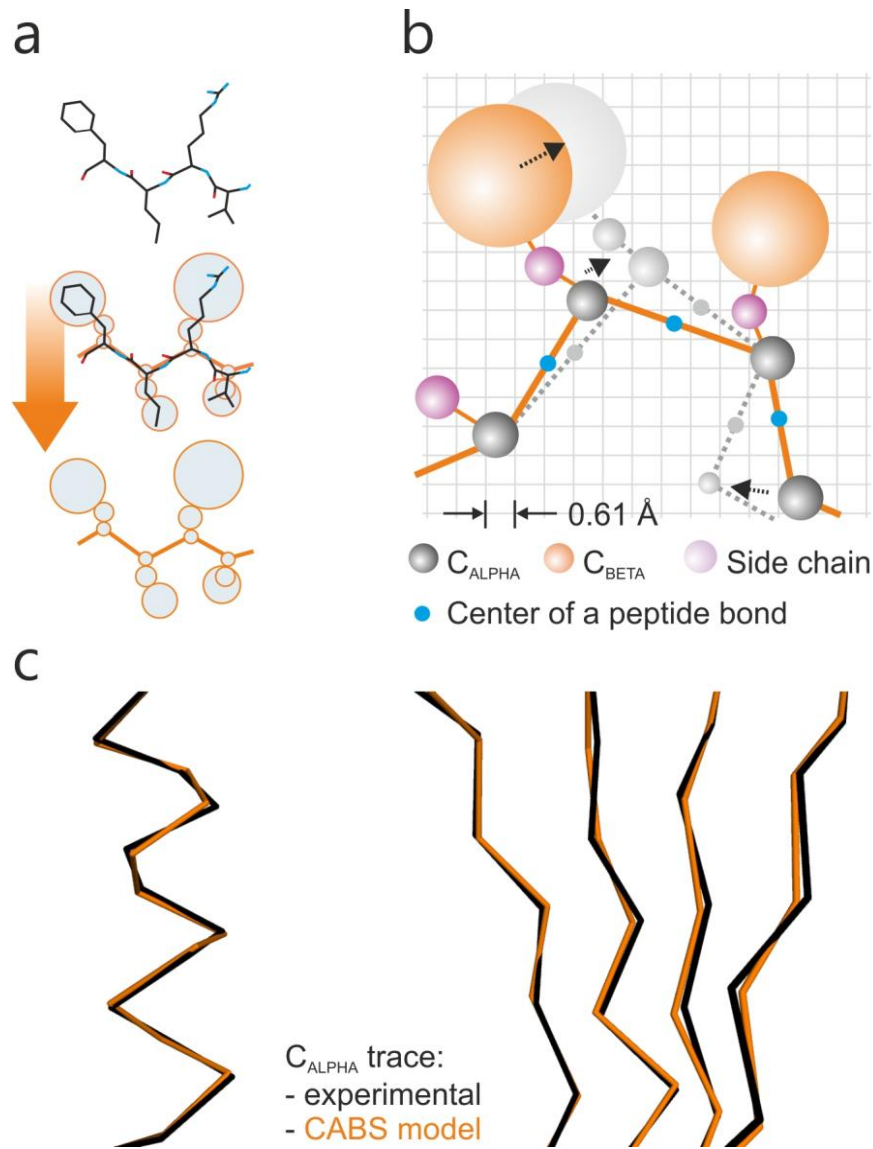

**Fig 1.** Representation of a protein chain in the CABS model. (a) scheme showing conversion from all-atom to coarse-grained CABS representation, (b) details of CABS coarse-grained representation, (c) comparison of the C-alpha trace in experimental protein structure (black color) and after conversion to CABS representation (orange) presented here for an example helix and beta sheet secondary structure (experimentally derived C-alpha coordinates of both secondary structure motifs were taken from the 2GB1 PDB file).

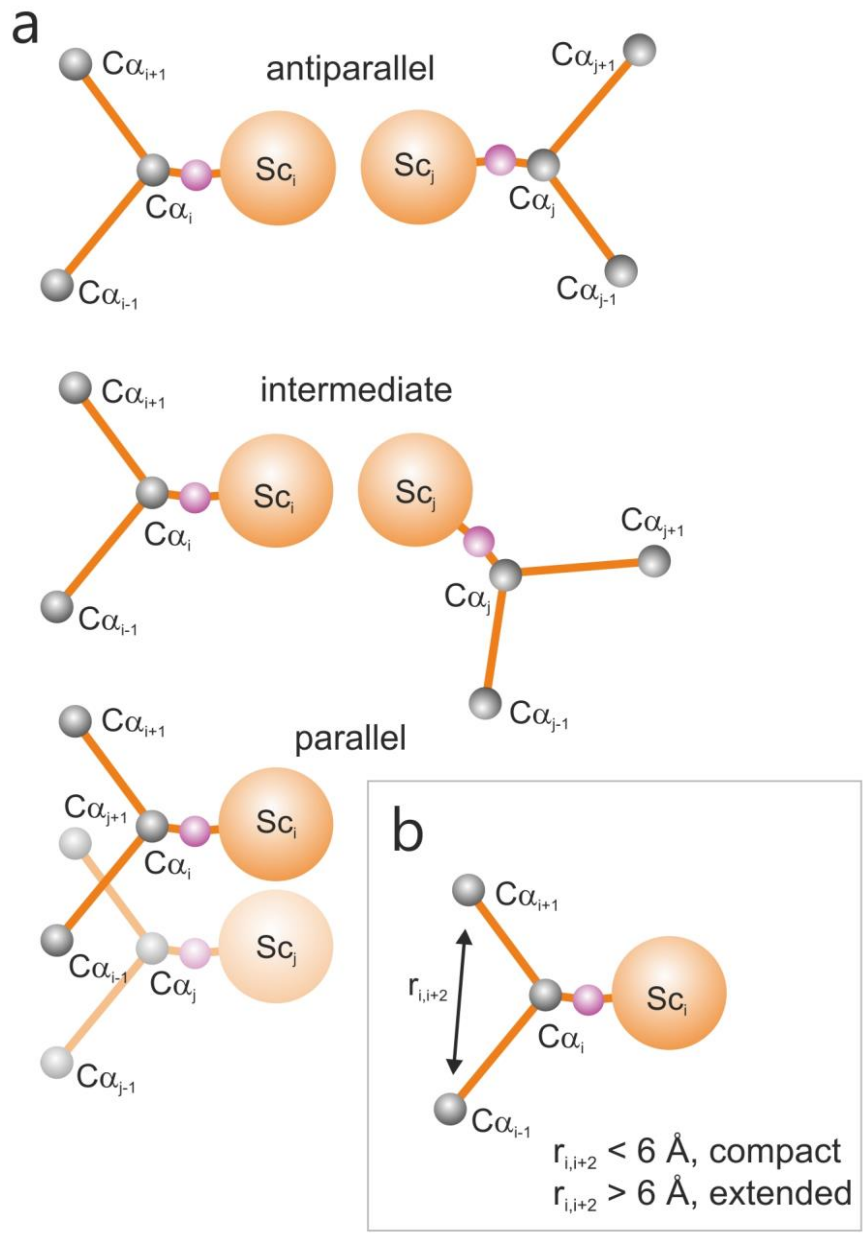

**Fig 2.** Types of protein structure arrangements used in the definition of sequence-dependent pairwise potentials. (a) Three types of mutual orientations of the side chains (antiparallel, medium-intermediate, parallel). (b) Two types of main chain conformations (helical-compact and expanded-beta). Numerical values of the potentials are given in Tables 2 to 10.

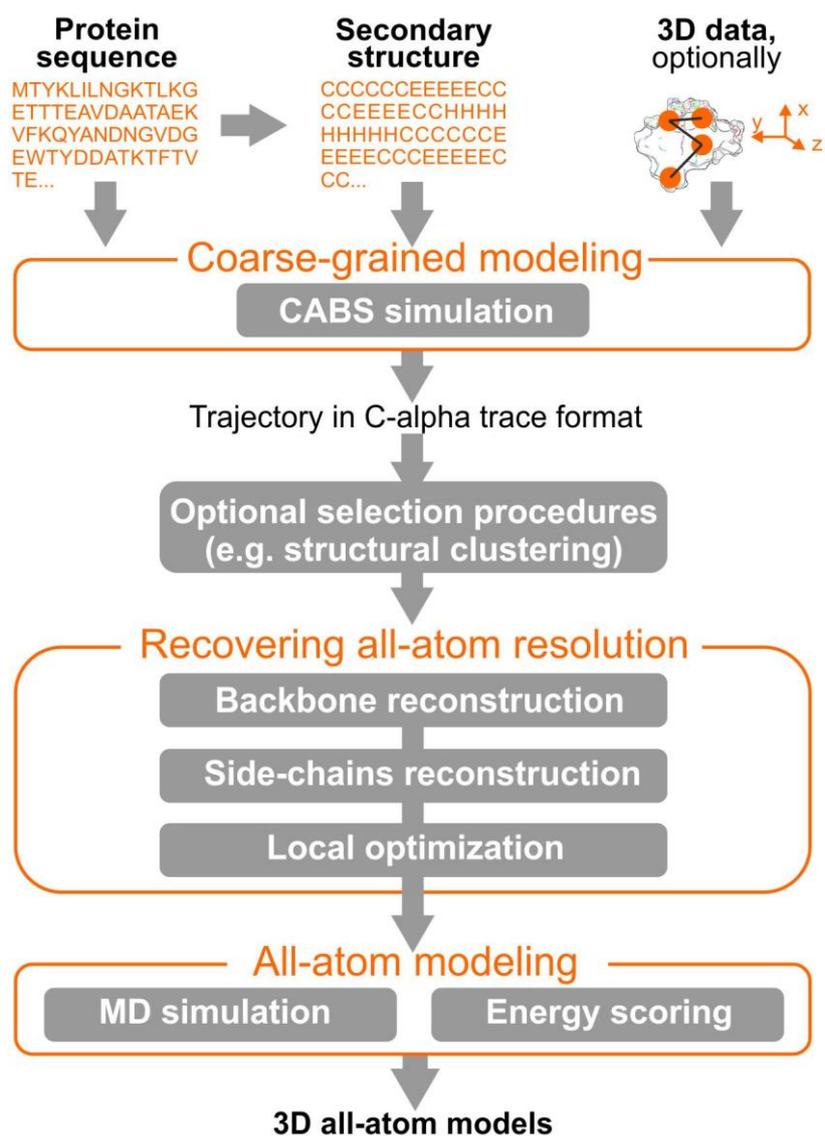

**Fig 3.** Typical stages of the multiscale modeling scheme utilizing the CABS model. The modeling input includes one dimensional data (protein sequence and secondary structure) and, optionally, three-dimensional data (e.g. distance restraints from experiment or from evolutionary analysis). Secondary structure data are required in a three letter code (C, coil; E, extended; H, helix). The modeling scheme consists of three major stages: (1) coarse-grained modeling with the CABS model, (2) several steps of reconstruction to all-atom representation and (3) all-atom modeling procedures (e.g. simulation using all-atom MD or all-atom energy scoring).

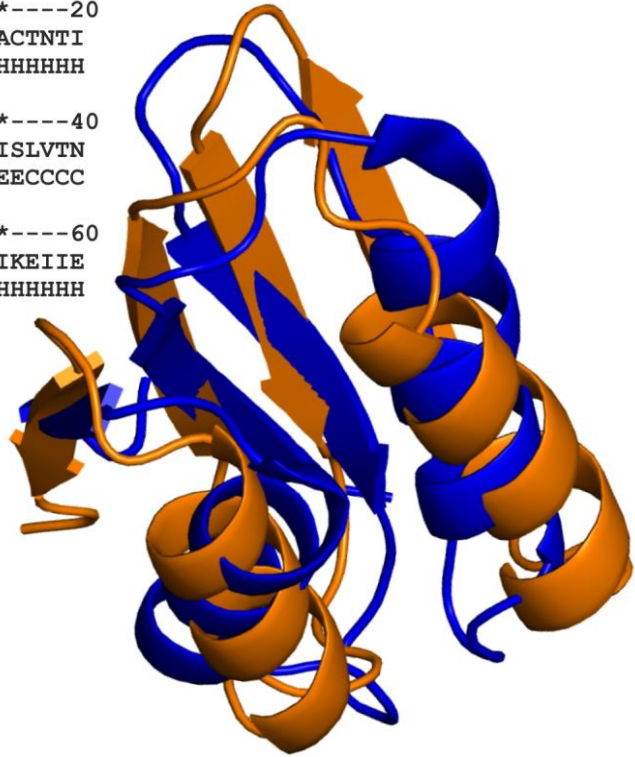

**Fig 4.** Example screenshots from the CABS-fold server. (a) Main page input panel. Output panels presenting: (b) predicted models, (c) RMSD between the predicted models, (d) characteristics of the structure prediction trajectories. Selected/clicked options are marked with orange rectangles and arrows.

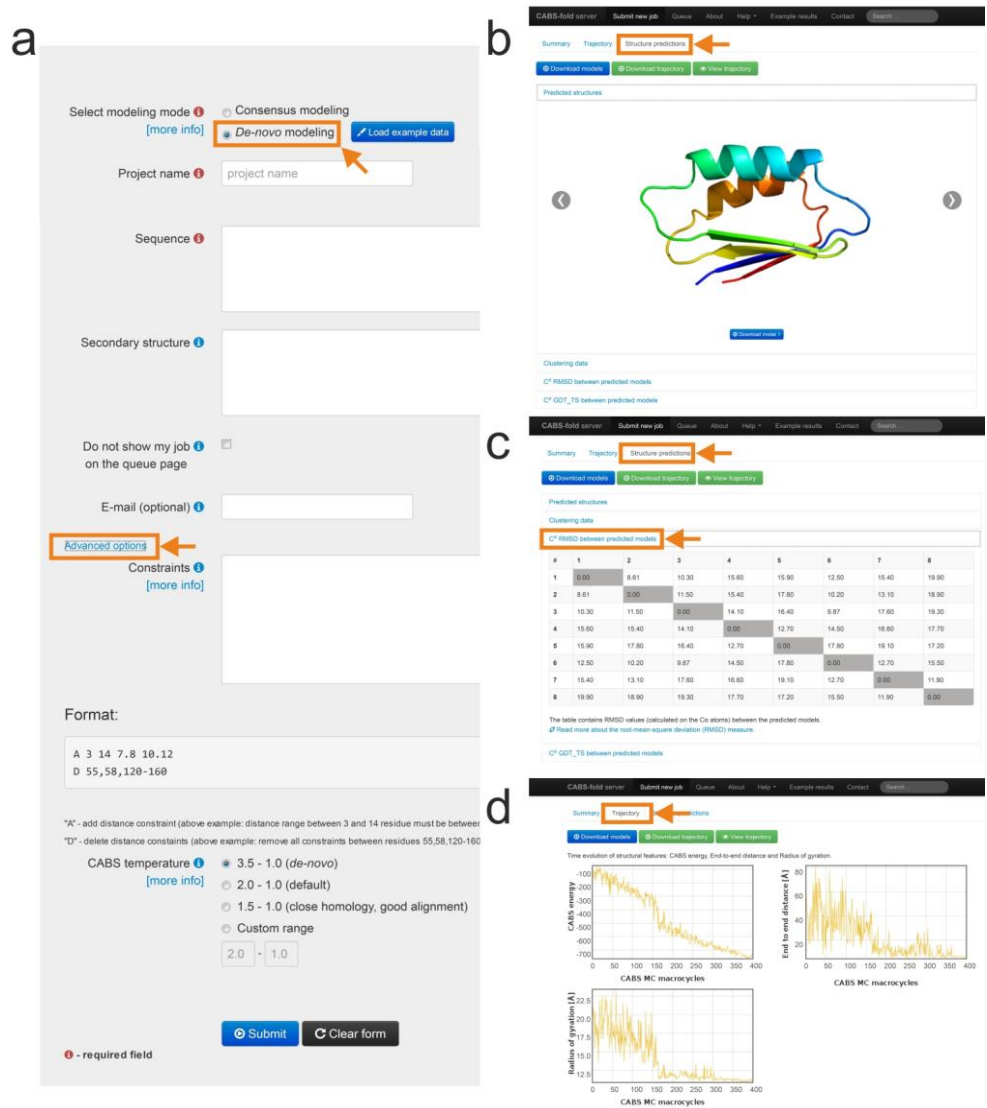

**Fig 5.** Example CABS-fold structure prediction result. The only input data: protein sequence and secondary structure (predicted from sequence by the Psi-pred method [ref]) are shown on the left. The experimental structure (blue) of a 72 residue protein (PDB ID: 1fvq) is superimposed on the CABS-fold predicted model (orange). In comparison to the experimental structure, the CABS-fold model has the same fold and RMSD value is 3.7 Å.

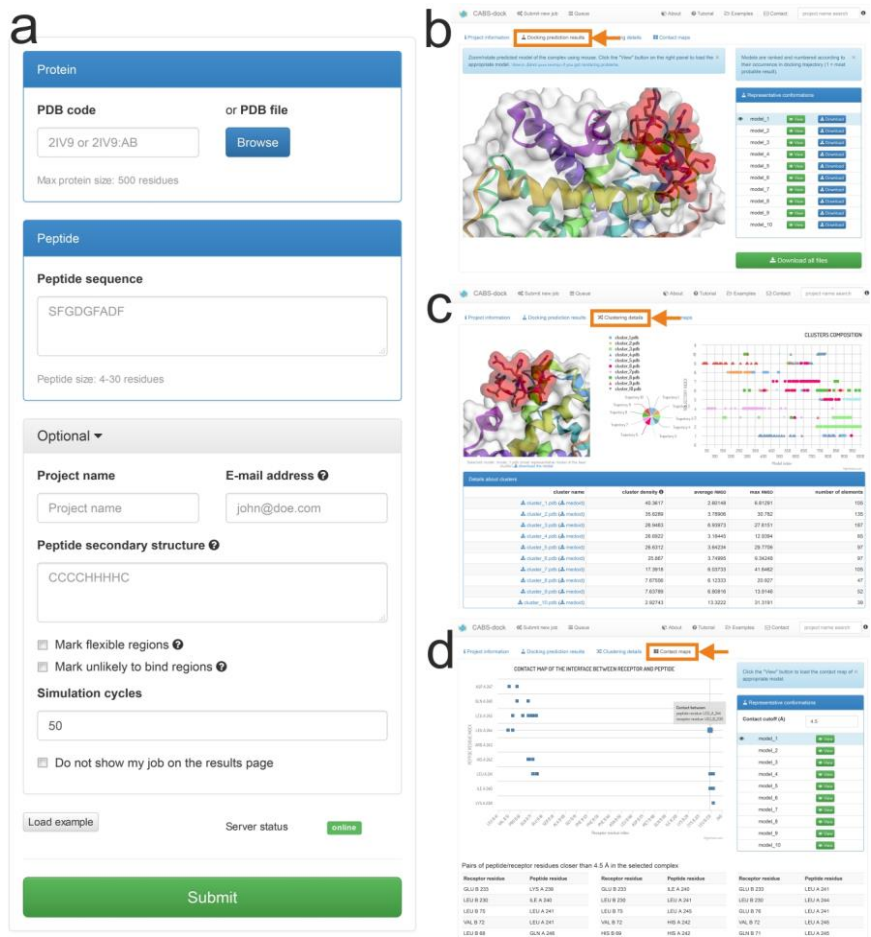

**Fig 6.** Example screenshots from the CABS-dock server. (a) Main page input panel. Output panels presenting: (b) predicted models, (c) clustering results and analysis, (d) contact maps for predicted models. Selected/clicked options are marked with orange rectangles and arrows.

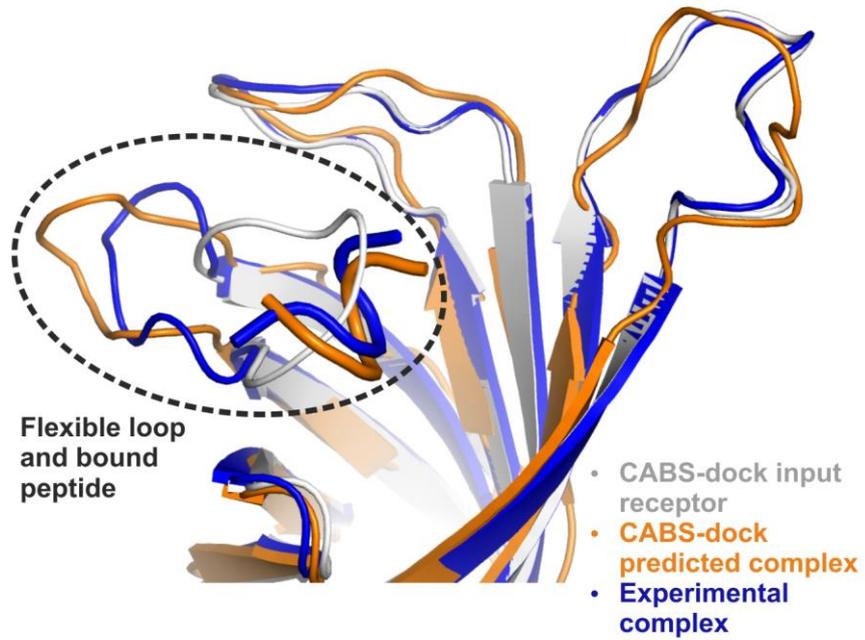

**Fig 7.** Example result of CABS-dock protein-peptide docking using the option of significant flexibility for the selected receptor fragment. The figure shows comparison of the CABS-dock input structure in the peptide-unbound form (colored in gray, PDB ID: 2RTM) with a CABS-dock-predicted complex (in orange) and a peptide-bound experimental complex (in blue, PDB ID: 1KL3). RMSD between the predicted and experimental peptide structure is 2.03 Å. The flexible loop region (designated to be fully flexible during docking) is between residues 45 and 54.

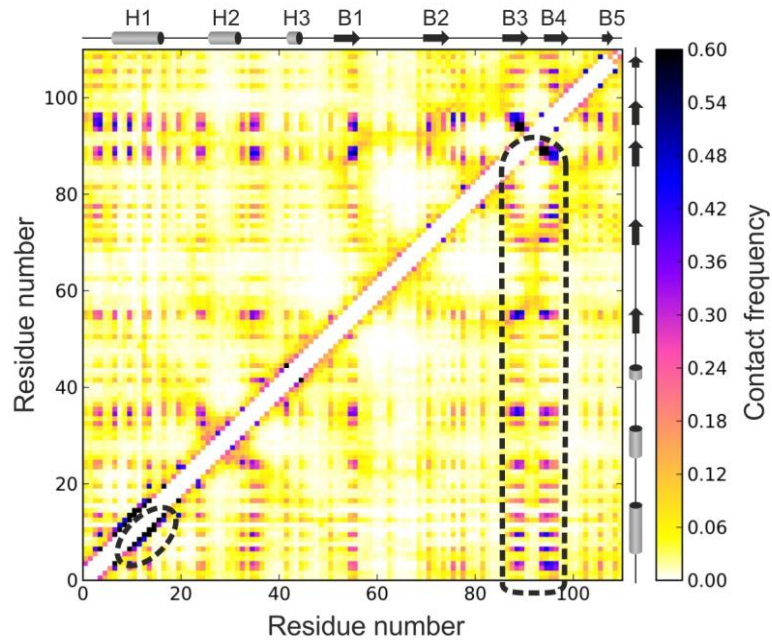

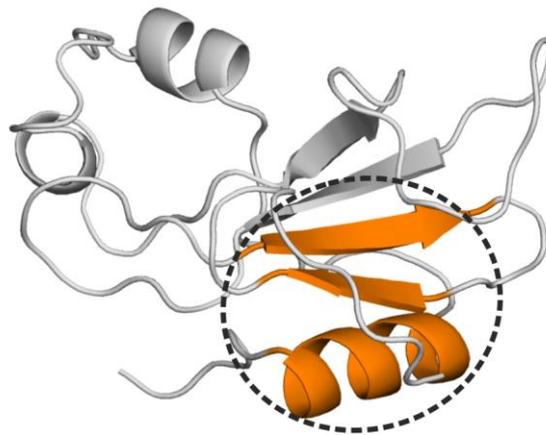

**Fig 8.** Example result from simulations of long-term protein dynamics (from a fully denatured to a near-native state) using the CABS model and protein sequence only. A simulation contact map is presented showing the key step of barnase folding (PDB code: 1BNR). The presented key folding step is the formation of the nucleation site. The nucleation site is formed by the following elements of secondary structure: helix 1 and beta-strands: 3 (marked by dashed lines in the contact map and colored also in orange in the native barnase structure, shown below). The map colors indicate contact frequency (see the legend).

**Table captions (from 2 to 10)**

**Table 2.** Context pairwise contact potential for the PCC type of side chain interactions. The PCC type of interactions occurs between the amino acid chains forming a parallel contact (P), where both contacting side chains are attached to a compact (C, most likely a helix) type of conformation.

**Table 3.** Context pairwise contact potential for the PEE type of side chain interactions. The PEE type of interactions occurs between the amino acid chains forming a parallel contact (P), where both contacting side chains are attached to expanded (E, most likely beta-strand) type of conformation.

**Table 4.** Context pairwise contact potential for the PCE and the PEC type of side chain interactions. The PCE type of interactions occurs between amino acid chains forming a parallel contact (P), where the first contacting side chain (given in columns) is attached to the compact (C, most likely a helix) type of conformation and the second contacting side chain (given in rows) is attached to expanded (E, most likely beta-strand) conformation.

**Table 5.** Context pairwise contact potential for the MCC type of side chain interactions. The MCC type of interactions occurs between the amino acid chains forming a medium-intermediate contact (M), where both contacting side chains are attached to a compact (C, most likely a helix) type of conformation.

**Table 6.** Context pairwise contact potential for the MEE type of side chain interactions. The MEE type of interactions occurs between the amino acid chains forming a medium-intermediate contact (M), where both contacting side chains are attached to expanded (E, most likely beta-strand) type of conformation.

**Table 7.** Context pairwise contact potential for the MCE and the MEC type of side chain interactions. The MCE type of interactions occurs between amino acid chains forming a medium-intermediate contact (M), where the first contacting side chain (given in columns) is attached to the compact (C, most likely a helix) type of conformation and the second contacting side chain (given in rows) is attached to expanded (E, most likely beta-strand) conformation.

**Table 8.** Context pairwise contact potential for the ACC type of side chain interactions. The ACC type of interactions occurs between the amino acid chains forming an antiparallel contact (A), where both contacting side chains are attached to a compact (C, most likely a helix) type of conformation.

**Table 9.** Context pairwise contact potential for the AEE type of side chain interactions. The AEE type of interactions occurs between the amino acid chains forming an antiparallel contact (A), where both contacting side chains are attached to expanded (E, most likely beta-strand) type of conformation.

**Table 10.** Context pairwise contact potential for the ACE and the AEC type of side chain interactions. The ACE type of interactions occurs between amino acid chains forming an antiparallel contact (A), where the first contacting side chain (given in columns) is attached to the compact (C, most likely a helix) type of conformation and the second contacting side chain (given in rows) is attached to expanded (E, most likely beta-strand) conformation.

**Table 2**

| PCC | GLY | ALA | SER | CYS | VAL | THR | ILE | PRO | MET | ASP | ASN | LEU | LYS | GLU | GLN | ARG | HIS | PHE | TYR | TRP |
|---|---|---|---|---|---|---|---|---|---|---|---|---|---|---|---|---|---|---|---|---|
| **GLY** | 3.3 | 1.8 | 1.6 | 1.6 | 1.6 | 1.2 | 1.3 | 1.8 | 0.9 | 1.0 | 0.9 | 1.6 | 1.5 | 1.4 | 1.4 | 0.9 | 1.5 | 1.1 | 1.1 | 1.0 |
| **ALA** | 1.8 | 0.6 | 0.8 | 0.4 | 0.4 | 0.5 | 0.3 | 0.9 | 0.2 | 0.7 | 0.5 | 0.3 | 0.8 | 0.5 | 0.3 | 0.2 | 0.4 | 0.1 | 0.2 | 0.2 |
| **SER** | 1.6 | 0.8 | 0.7 | 0.9 | 0.3 | 0.4 | 0.2 | 0.9 | 0.3 | 0.2 | 0.2 | 0.5 | 0.3 | 0.2 | 0.0 | -0.1 | -0.1 | 0.1 | 0.0 | 0.0 |
| **CYS** | 1.6 | 0.4 | 0.9 | -0.2 | -0.4 | 0.3 | -0.6 | 0.4 | -0.6 | 0.9 | 0.4 | -0.4 | 0.3 | 0.5 | 0.7 | 0.2 | -0.4 | -0.8 | -0.6 | -0.6 |
| **VAL** | 1.6 | 0.4 | 0.3 | -0.4 | -0.6 | -0.3 | -0.9 | 0.6 | -0.8 | 0.7 | -0.0 | -0.9 | 0.1 | 0.2 | -0.0 | -0.3 | -0.3 | -1.0 | -0.7 | -1.1 |
| **THR** | 1.2 | 0.5 | 0.4 | 0.3 | -0.3 | -0.1 | -0.5 | 0.7 | -0.2 | 0.1 | -0.2 | -0.3 | -0.1 | -0.0 | -0.4 | -0.2 | -0.2 | -0.3 | -0.4 | -0.4 |
| **ILE** | 1.3 | 0.3 | 0.2 | -0.6 | -0.9 | -0.5 | -1.1 | 0.4 | -0.9 | 0.5 | 0.1 | -1.1 | -0.1 | -0.0 | -0.2 | -0.3 | -0.3 | -1.1 | -0.7 | -1.1 |
| **PRO** | 1.8 | 0.9 | 0.9 | 0.4 | 0.6 | 0.7 | 0.4 | 1.4 | 0.4 | 1.0 | 0.6 | 0.6 | 0.6 | 0.6 | 0.5 | 0.2 | 0.5 | 0.6 | -0.0 | -0.0 |
| **MET** | 0.9 | 0.2 | 0.3 | -0.6 | -0.8 | -0.2 | -0.9 | 0.4 | -0.2 | 0.4 | -0.1 | -0.9 | 0.1 | 0.1 | -0.3 | -0.5 | -0.7 | -1.1 | -0.9 | -0.8 |
| **ASP** | 1.0 | 0.7 | 0.2 | 0.9 | 0.7 | 0.1 | 0.5 | 1.0 | 0.4 | 0.5 | 0.0 | 0.9 | -0.6 | 0.5 | -0.4 | -0.7 | -0.2 | 0.3 | 0.3 | 0.2 |
| **ASN** | 0.9 | 0.5 | 0.2 | 0.4 | -0.0 | -0.2 | 0.1 | 0.6 | -0.1 | 0.0 | 0.1 | 0.2 | -0.3 | -0.1 | -0.3 | -0.3 | -0.0 | 0.1 | -0.1 | -0.1 |
| **LEU** | 1.6 | 0.3 | 0.5 | -0.4 | -0.9 | -0.3 | -1.1 | 0.6 | -0.9 | 0.9 | 0.2 | -1.0 | -0.0 | 0.2 | -0.1 | -0.3 | -0.3 | -1.1 | -0.8 | -0.9 |
| **LYS** | 1.5 | 0.8 | 0.3 | 0.3 | 0.1 | -0.1 | -0.1 | 0.6 | 0.1 | -0.6 | -0.3 | -0.0 | 0.4 | -0.9 | -0.5 | -0.1 | 0.1 | -0.0 | -0.4 | 0.0 |
| **GLU** | 1.4 | 0.5 | 0.2 | 0.5 | 0.2 | -0.0 | -0.0 | 0.6 | 0.1 | 0.5 | -0.1 | 0.2 | -0.9 | 0.2 | -0.5 | -1.0 | -0.4 | 0.2 | -0.1 | -0.2 |
| **GLN** | 1.4 | 0.3 | 0.0 | 0.7 | -0.0 | -0.4 | -0.2 | 0.5 | -0.3 | -0.4 | -0.3 | -0.1 | -0.5 | -0.5 | -0.2 | -0.5 | -0.3 | -0.1 | -0.4 | -0.7 |
| **ARG** | 0.9 | 0.2 | -0.1 | 0.2 | -0.3 | -0.2 | -0.3 | 0.2 | -0.5 | -0.7 | -0.3 | -0.3 | -0.1 | -1.0 | -0.5 | -0.2 | -0.3 | -0.3 | -0.6 | -0.5 |
| **HIS** | 1.5 | 0.4 | -0.1 | -0.4 | -0.3 | -0.2 | -0.3 | 0.5 | -0.7 | -0.2 | -0.0 | -0.3 | 0.1 | -0.4 | -0.3 | -0.3 | -0.4 | -0.5 | -0.7 | -0.8 |
| **PHE** | 1.1 | 0.1 | 0.1 | -0.8 | -1.0 | -0.3 | -1.1 | 0.6 | -1.1 | 0.3 | 0.1 | -1.1 | -0.0 | 0.2 | -0.1 | -0.3 | -0.5 | -1.0 | -1.0 | -1.2 |
| **TYR** | 1.1 | 0.2 | 0.0 | -0.6 | -0.7 | -0.4 | -0.7 | -0.0 | -0.9 | 0.3 | -0.1 | -0.8 | -0.4 | -0.1 | -0.4 | -0.6 | -0.7 | -1.0 | -0.5 | -1.1 |
| **TRP** | 1.0 | 0.2 | 0.0 | -0.6 | -1.1 | -0.4 | -1.1 | -0.0 | -0.8 | 0.2 | -0.1 | -0.9 | 0.0 | -0.2 | -0.7 | -0.5 | -0.8 | -1.2 | -1.1 | -1.1 |

**Table 3**

| PEE | GLY | ALA | SER | CYS | VAL | THR | ILE | PRO | MET | ASP | ASN | LEU | LYS | GLU | GLN | ARG | HIS | PHE | TYR | TRP |
|---|---|---|---|---|---|---|---|---|---|---|---|---|---|---|---|---|---|---|---|---|
| **GLY** | 2.6 | 1.3 | 0.9 | 0.8 | 0.8 | 0.8 | 0.6 | 1.4 | 0.5 | 0.5 | 0.8 | 0.5 | 1.0 | 0.8 | 0.6 | 0.6 | 0.3 | 0.0 | 0.1 | -0.2 |
| **ALA** | 1.3 | 0.5 | 0.5 | -0.0 | -0.4 | 0.2 | -0.6 | 0.6 | -0.2 | 0.6 | 0.3 | -0.6 | 0.5 | 0.6 | 0.3 | 0.1 | 0.1 | -0.8 | -0.6 | -0.8 |
| **SER** | 0.9 | 0.5 | 0.0 | 0.1 | 0.1 | -0.3 | -0.1 | 0.6 | -0.1 | -0.3 | -0.3 | 0.0 | -0.0 | -0.2 | -0.3 | -0.3 | -0.6 | -0.4 | -0.5 | -0.7 |
| **CYS** | 0.8 | -0.0 | 0.1 | -0.8 | -0.9 | -0.0 | -0.9 | 0.2 | -0.7 | 0.3 | 0.3 | -0.9 | 0.1 | 0.4 | -0.1 | 0.0 | -0.6 | -0.9 | -0.7 | -1.5 |
| **VAL** | 0.8 | -0.4 | 0.1 | -0.9 | -1.3 | -0.5 | -1.5 | 0.2 | -0.9 | 0.3 | 0.0 | -1.4 | -0.2 | -0.1 | -0.1 | -0.5 | -0.6 | -1.3 | -1.1 | -1.2 |
| **THR** | 0.8 | 0.2 | -0.3 | -0.0 | -0.5 | -0.7 | -0.6 | 0.3 | -0.4 | -0.2 | -0.5 | -0.4 | -0.5 | -0.6 | -0.6 | -0.7 | -0.7 | -0.5 | -0.6 | -0.6 |
| **ILE** | 0.6 | -0.6 | -0.1 | -0.9 | -1.5 | -0.6 | -1.6 | 0.1 | -1.1 | 0.1 | -0.0 | -1.5 | -0.2 | -0.2 | -0.2 | -0.5 | -0.5 | -1.5 | -1.1 | -1.3 |
| **PRO** | 1.4 | 0.6 | 0.6 | 0.2 | 0.2 | 0.3 | 0.1 | 1.0 | 0.2 | 0.8 | 0.4 | 0.1 | 0.8 | 0.6 | 0.3 | 0.1 | 0.0 | -0.3 | -0.5 | -0.5 |
| **MET** | 0.5 | -0.2 | -0.1 | -0.7 | -0.9 | -0.4 | -1.1 | 0.2 | -0.4 | 0.2 | -0.1 | -1.1 | -0.2 | -0.2 | -0.5 | -0.5 | -0.5 | -1.3 | -0.9 | -1.2 |
| **ASP** | 0.5 | 0.6 | -0.3 | 0.3 | 0.3 | -0.2 | 0.1 | 0.8 | 0.2 | 0.4 | -0.5 | 0.2 | -0.6 | 0.1 | -0.5 | -1.0 | -0.7 | 0.0 | -0.4 | -0.4 |
| **ASN** | 0.8 | 0.3 | -0.3 | 0.3 | 0.0 | -0.5 | -0.0 | 0.4 | -0.1 | -0.5 | -0.5 | 0.1 | -0.3 | -0.4 | -0.6 | -0.5 | -0.4 | -0.3 | -0.6 | -0.6 |
| **LEU** | 0.5 | -0.6 | 0.0 | -0.9 | -1.4 | -0.4 | -1.5 | 0.1 | -1.1 | 0.2 | 0.1 | -1.3 | -0.1 | -0.2 | -0.3 | -0.4 | -0.5 | -1.4 | -1.1 | -1.2 |
| **LYS** | 1.0 | 0.5 | -0.0 | 0.1 | -0.2 | -0.5 | -0.2 | 0.8 | -0.2 | -0.6 | -0.3 | -0.1 | 0.3 | -1.1 | -0.6 | -0.1 | -0.3 | -0.3 | -0.8 | -1.0 |
| **GLU** | 0.8 | 0.6 | -0.2 | 0.4 | -0.1 | -0.6 | -0.2 | 0.6 | -0.2 | 0.1 | -0.4 | -0.2 | -1.1 | 0.4 | -0.3 | -1.3 | -0.8 | -0.3 | -0.8 | -0.7 |
| **GLN** | 0.6 | 0.3 | -0.3 | -0.1 | -0.1 | -0.6 | -0.2 | 0.3 | -0.5 | -0.5 | -0.6 | -0.3 | -0.6 | -0.3 | -0.3 | -0.8 | -0.6 | -0.6 | -0.8 | -1.0 |
| **ARG** | 0.6 | 0.1 | -0.3 | 0.0 | -0.5 | -0.7 | -0.5 | 0.1 | -0.5 | -1.0 | -0.5 | -0.4 | -0.1 | -1.3 | -0.8 | -0.4 | -0.7 | -0.7 | -1.0 | -0.8 |
| **HIS** | 0.3 | 0.1 | -0.6 | -0.6 | -0.6 | -0.7 | -0.5 | 0.0 | -0.5 | -0.7 | -0.4 | -0.5 | -0.3 | -0.8 | -0.6 | -0.7 | -0.6 | -0.6 | -1.0 | -0.7 |
| **PHE** | 0.0 | -0.8 | -0.4 | -0.9 | -1.3 | -0.5 | -1.5 | -0.3 | -1.3 | 0.0 | -0.3 | -1.4 | -0.3 | -0.3 | -0.6 | -0.7 | -0.6 | -1.4 | -1.2 | -1.4 |
| **TYR** | 0.1 | -0.6 | -0.5 | -0.7 | -1.1 | -0.6 | -1.1 | -0.5 | -0.9 | -0.4 | -0.6 | -1.1 | -0.8 | -0.8 | -0.8 | -1.0 | -1.0 | -1.2 | -0.9 | -1.1 |
| **TRP** | -0.2 | -0.8 | -0.7 | -1.5 | -1.2 | -0.6 | -1.3 | -0.5 | -1.2 | -0.4 | -0.6 | -1.2 | -1.0 | -0.7 | -1.0 | -0.8 | -0.7 | -1.4 | -1.1 | -0.9 |

**Table 4**

| PCE/PEC | GLY | ALA | SER | CYS | VAL | THR | ILE | PRO | MET | ASP | ASN | LEU | LYS | GLU | GLN | ARG | HIS | PHE | TYR | TRP |
|---|---|---|---|---|---|---|---|---|---|---|---|---|---|---|---|---|---|---|---|---|
| **GLY** | 3.3 | 2.5 | 1.4 | 1.5 | 1.3 | 0.9 | 1.2 | 2.2 | 0.8 | 0.8 | 1.1 | 1.3 | 0.9 | 1.0 | 0.9 | 0.7 | 1.1 | 1.0 | 0.6 | 0.6 |
| **ALA** | 2.3 | 1.4 | 0.8 | 0.7 | 1.1 | 0.9 | 1.0 | 1.5 | 0.8 | 0.4 | 0.3 | 1.0 | 1.3 | 1.3 | 1.2 | 0.9 | 0.9 | 0.6 | 0.5 | 0.4 |
| **SER** | 1.3 | 0.8 | 0.4 | 0.6 | 1.1 | 0.5 | 1.0 | 0.5 | 0.9 | -0.5 | 0.3 | 0.9 | 0.5 | 0.4 | 0.2 | 0.1 | 0.5 | 0.6 | 0.6 | 0.5 |
| **CYS** | 0.8 | 0.6 | 0.5 | -1.2 | 0.2 | 0.3 | -0.1 | 0.7 | 0.0 | 0.8 | 0.2 | 0.0 | 0.6 | 1.0 | 0.5 | -0.1 | -0.2 | -0.3 | -0.0 | -0.8 |
| **VAL** | 1.6 | 0.6 | 0.6 | 0.3 | 0.1 | 0.5 | -0.0 | 0.7 | 0.0 | 0.1 | 0.2 | -0.2 | 0.5 | 0.2 | 0.4 | 0.3 | 0.2 | -0.0 | 0.0 | 0.2 |
| **THR** | 1.3 | 0.9 | 0.3 | 0.8 | 0.8 | 0.3 | 0.6 | 0.7 | 0.5 | -0.3 | 0.0 | 0.6 | 0.1 | 0.1 | 0.3 | 0.1 | 0.2 | 0.5 | 0.1 | 0.4 |
| **ILE** | 1.3 | 0.5 | 0.4 | 0.2 | 0.1 | 0.3 | -0.2 | 0.6 | -0.2 | 0.6 | 0.2 | -0.3 | 0.7 | 0.5 | 0.3 | 0.5 | 0.5 | -0.4 | -0.1 | -0.2 |
| **PRO** | 1.6 | 1.4 | 1.3 | 1.4 | 0.9 | 1.0 | 0.8 | 1.2 | 1.1 | 1.0 | 0.4 | 0.8 | 1.0 | 0.7 | 0.6 | 0.7 | 0.8 | 0.3 | -0.1 | -0.1 |
| **MET** | 1.0 | 0.4 | 0.5 | 0.0 | 0.3 | 0.8 | 0.0 | 0.5 | -0.1 | 0.4 | 0.1 | -0.0 | 0.6 | 1.0 | 0.3 | 0.2 | -0.1 | -0.2 | 0.0 | -0.4 |
| **ASP** | 1.0 | 1.1 | 0.1 | 1.1 | 1.2 | 0.1 | 1.2 | 0.8 | 0.8 | 0.2 | -0.2 | 1.0 | -0.2 | 0.8 | 0.3 | -0.5 | -0.1 | 1.0 | 0.2 | 0.6 |
| **ASN** | 1.3 | 1.2 | 0.5 | 1.2 | 0.9 | 0.3 | 1.1 | 0.9 | 0.7 | -0.2 | -0.2 | 0.8 | 0.3 | 0.3 | 0.2 | 0.0 | 0.2 | 0.9 | 0.1 | 0.2 |
| **LEU** | 1.3 | 0.5 | 0.6 | 0.1 | -0.0 | 0.6 | -0.3 | 0.8 | -0.0 | 0.6 | 0.4 | -0.4 | 0.7 | 0.7 | 0.5 | 0.3 | 0.1 | -0.3 | -0.2 | -0.4 |
| **LYS** | 1.8 | 1.9 | 0.6 | 1.6 | 1.3 | 0.4 | 1.3 | 1.0 | 0.8 | -0.2 | 0.2 | 1.1 | 1.1 | -0.1 | 0.5 | 0.7 | 0.5 | 0.8 | 0.4 | 0.4 |
| **GLU** | 1.3 | 1.3 | 0.1 | 1.1 | 1.6 | 0.2 | 1.3 | 0.7 | 0.9 | 0.5 | 0.1 | 1.3 | -0.1 | 0.9 | 0.6 | -0.3 | 0.2 | 1.0 | 0.7 | 0.4 |
| **GLN** | 1.0 | 1.1 | 0.4 | 1.3 | 1.4 | 0.5 | 0.8 | 0.7 | 0.6 | 0.1 | 0.3 | 0.8 | 0.4 | 0.5 | 0.3 | 0.1 | 0.6 | 0.6 | 0.5 | -0.1 |
| **ARG** | 1.1 | 1.2 | 0.5 | 0.6 | 1.0 | 0.5 | 0.8 | 0.4 | 1.0 | -0.5 | 0.1 | 0.7 | 0.5 | -0.1 | 0.1 | 0.4 | 0.2 | 0.5 | 0.1 | 0.4 |
| **HIS** | 1.3 | 0.8 | 0.2 | -0.1 | 0.6 | 0.2 | 0.5 | 0.3 | 0.6 | 0.0 | -0.0 | 0.6 | 0.1 | 0.0 | 0.1 | 0.2 | -0.1 | 0.1 | -0.2 | -0.4 |
| **PHE** | 1.0 | 0.2 | 0.2 | -0.3 | -0.2 | 0.1 | -0.4 | 0.6 | -0.5 | 0.3 | 0.0 | -0.4 | 0.6 | 0.4 | 0.3 | 0.1 | -0.1 | -0.7 | -0.4 | -0.3 |
| **TYR** | 1.0 | 0.6 | 0.2 | 0.2 | 0.2 | 0.1 | -0.0 | -0.1 | -0.2 | 0.0 | -0.2 | -0.1 | 0.2 | 0.2 | 0.3 | -0.1 | -0.4 | -0.4 | -0.1 | -0.5 |
| **TRP** | 0.7 | 0.0 | 0.0 | -0.1 | -0.0 | -0.2 | -0.5 | -0.4 | -0.6 | 0.2 | -0.1 | -0.3 | 0.2 | -0.2 | -0.3 | 0.1 | -0.3 | -0.7 | -0.4 | -0.4 |

**Table 5**

| MCC | GLY | ALA | SER | CYS | VAL | THR | ILE | PRO | MET | ASP | ASN | LEU | LYS | GLU | GLN | ARG | HIS | PHE | TYR | TRP |
|---|---|---|---|---|---|---|---|---|---|---|---|---|---|---|---|---|---|---|---|---|
| **GLY** | 2.0 | 1.4 | 1.1 | 0.7 | 1.0 | 1.0 | 0.7 | 1.2 | 0.5 | 0.9 | 0.9 | 0.7 | 1.0 | 1.3 | 0.8 | 0.5 | 1.0 | 0.4 | 0.4 | 0.2 |
| **ALA** | 1.4 | 0.3 | 0.4 | 0.0 | -0.2 | 0.0 | -0.3 | 0.5 | -0.3 | 0.5 | 0.5 | -0.3 | 0.4 | 0.3 | 0.3 | 0.1 | 0.4 | -0.4 | -0.4 | -0.5 |
| **SER** | 1.1 | 0.4 | 0.6 | 0.4 | 0.1 | 0.2 | 0.2 | 0.5 | 0.2 | 0.3 | 0.4 | 0.1 | 0.2 | 0.5 | 0.1 | 0.1 | 0.1 | -0.1 | -0.1 | -0.2 |
| **CYS** | 0.7 | 0.0 | 0.4 | -1.3 | -0.3 | 0.2 | -0.5 | 0.3 | -0.7 | 0.6 | 0.5 | -0.5 | 0.5 | 0.5 | 0.8 | 0.1 | -0.6 | -0.9 | -0.6 | -0.9 |
| **VAL** | 1.0 | -0.2 | 0.1 | -0.3 | -0.5 | -0.1 | -0.8 | -0.0 | -0.8 | 0.5 | 0.5 | -0.8 | 0.5 | 0.2 | 0.1 | 0.1 | -0.1 | -0.8 | -0.7 | -1.1 |
| **THR** | 1.0 | 0.0 | 0.2 | 0.2 | -0.1 | 0.2 | -0.2 | 0.2 | -0.3 | 0.3 | 0.0 | -0.2 | 0.2 | 0.3 | 0.0 | -0.0 | 0.0 | -0.3 | -0.3 | -0.6 |
| **ILE** | 0.7 | -0.3 | 0.2 | -0.5 | -0.8 | -0.2 | -0.8 | 0.1 | -1.0 | 0.5 | 0.3 | -1.0 | 0.2 | 0.1 | 0.0 | -0.2 | -0.2 | -1.1 | -0.9 | -1.2 |
| **PRO** | 1.2 | 0.5 | 0.5 | 0.3 | -0.0 | 0.2 | 0.1 | 0.7 | 0.0 | 0.6 | 0.5 | 0.2 | 0.5 | 0.7 | 0.3 | -0.0 | 0.2 | -0.3 | -0.3 | -0.3 |
| **MET** | 0.5 | -0.3 | 0.2 | -0.7 | -0.8 | -0.3 | -1.0 | 0.0 | -0.6 | 0.4 | 0.3 | -0.9 | 0.2 | 0.3 | -0.1 | -0.3 | -0.4 | -1.1 | -0.9 | -1.2 |
| **ASP** | 0.9 | 0.5 | 0.3 | 0.6 | 0.5 | 0.3 | 0.5 | 0.6 | 0.4 | 0.5 | 0.2 | 0.6 | -0.3 | 0.4 | 0.4 | -0.6 | -0.1 | 0.3 | -0.2 | -0.1 |
| **ASN** | 0.9 | 0.5 | 0.4 | 0.5 | 0.5 | 0.0 | 0.3 | 0.5 | 0.3 | 0.2 | 0.3 | 0.4 | 0.1 | 0.2 | 0.2 | -0.1 | 0.1 | 0.1 | -0.3 | -0.4 |
| **LEU** | 0.7 | -0.3 | 0.1 | -0.5 | -0.8 | -0.2 | -1.0 | 0.2 | -0.9 | 0.6 | 0.4 | -1.1 | 0.2 | 0.2 | -0.1 | -0.1 | -0.2 | -1.2 | -0.9 | -1.3 |
| **LYS** | 1.0 | 0.4 | 0.2 | 0.5 | 0.5 | 0.2 | 0.2 | 0.5 | 0.2 | -0.3 | 0.1 | 0.2 | 0.9 | -0.4 | 0.0 | 0.3 | 0.4 | 0.1 | -0.3 | -0.1 |
| **GLU** | 1.3 | 0.3 | 0.5 | 0.5 | 0.2 | 0.3 | 0.1 | 0.7 | 0.3 | 0.4 | 0.2 | 0.2 | -0.4 | 0.4 | 0.0 | -0.6 | -0.2 | 0.0 | -0.4 | -0.2 |
| **GLN** | 0.8 | 0.3 | 0.1 | 0.8 | 0.1 | 0.0 | 0.0 | 0.3 | -0.1 | 0.4 | 0.2 | -0.1 | 0.0 | 0.0 | 0.1 | -0.1 | 0.0 | -0.1 | -0.3 | -0.5 |
| **ARG** | 0.5 | 0.1 | 0.1 | 0.1 | 0.1 | -0.0 | -0.2 | -0.0 | -0.3 | -0.6 | -0.1 | -0.1 | 0.3 | -0.6 | -0.1 | 0.1 | -0.0 | -0.3 | -0.6 | -0.6 |
| **HIS** | 1.0 | 0.4 | 0.1 | -0.6 | -0.1 | 0.0 | -0.2 | 0.2 | -0.4 | -0.1 | 0.1 | -0.2 | 0.4 | -0.2 | 0.0 | -0.0 | -0.0 | -0.3 | -0.3 | -1.0 |
| **PHE** | 0.4 | -0.4 | -0.1 | -0.9 | -0.8 | -0.3 | -1.1 | -0.3 | -1.1 | 0.3 | 0.1 | -1.2 | 0.1 | 0.0 | -0.1 | -0.3 | -0.3 | -1.2 | -1.2 | -1.5 |
| **TYR** | 0.4 | -0.4 | -0.1 | -0.6 | -0.7 | -0.3 | -0.9 | -0.3 | -0.9 | -0.2 | -0.3 | -0.9 | -0.3 | -0.4 | -0.3 | -0.6 | -0.3 | -1.2 | -0.7 | -1.3 |
| **TRP** | 0.2 | -0.5 | -0.2 | -0.9 | -1.1 | -0.6 | -1.2 | -0.3 | -1.2 | -0.1 | -0.4 | -1.3 | -0.1 | -0.2 | -0.5 | -0.6 | -1.0 | -1.5 | -1.3 | -1.2 |

**Table 6**

| MEE | GLY | ALA | SER | CYS | VAL | THR | ILE | PRO | MET | ASP | ASN | LEU | LYS | GLU | GLN | ARG | HIS | PHE | TYR | TRP |
|---|---|---|---|---|---|---|---|---|---|---|---|---|---|---|---|---|---|---|---|---|
| **GLY** | 1.5 | 1.2 | 0.8 | 0.2 | 0.6 | 0.5 | 0.6 | 0.8 | 0.3 | 0.7 | 0.3 | 0.5 | 0.9 | 0.7 | 0.7 | 0.3 | 0.1 | -0.0 | -0.3 | -0.5 |
| **ALA** | 1.2 | 0.4 | 0.5 | -0.3 | -0.3 | 0.2 | -0.4 | 0.3 | -0.2 | 0.5 | 0.3 | -0.2 | 0.5 | 0.4 | 0.4 | 0.1 | 0.1 | -0.5 | -0.4 | -0.8 |
| **SER** | 0.8 | 0.5 | 0.4 | 0.0 | 0.1 | 0.1 | 0.1 | 0.4 | 0.2 | 0.1 | -0.1 | 0.2 | 0.4 | 0.1 | 0.1 | 0.0 | -0.2 | -0.1 | -0.3 | -0.4 |
| **CYS** | 0.2 | -0.3 | 0.0 | -1.4 | -0.5 | -0.2 | -0.5 | -0.1 | -0.5 | 0.3 | 0.2 | -0.6 | 0.3 | 0.6 | -0.0 | -0.1 | -0.5 | -0.8 | -0.8 | -1.0 |
| **VAL** | 0.6 | -0.3 | 0.1 | -0.5 | -0.4 | 0.1 | -0.6 | -0.1 | -0.5 | 0.5 | 0.3 | -0.7 | 0.2 | 0.4 | 0.0 | -0.0 | -0.1 | -0.9 | -0.7 | -1.1 |
| **THR** | 0.5 | 0.2 | 0.1 | -0.2 | 0.1 | 0.2 | -0.0 | 0.1 | 0.1 | -0.0 | 0.1 | 0.0 | 0.2 | 0.1 | 0.1 | -0.2 | -0.0 | -0.2 | -0.3 | -0.4 |
| **ILE** | 0.6 | -0.4 | 0.1 | -0.5 | -0.6 | -0.0 | -0.8 | -0.2 | -0.6 | 0.4 | 0.3 | -0.8 | 0.3 | 0.4 | 0.1 | 0.0 | -0.2 | -1.0 | -0.8 | -1.1 |
| **PRO** | 0.8 | 0.3 | 0.4 | -0.1 | -0.1 | 0.1 | -0.2 | 0.5 | -0.3 | 0.6 | 0.2 | -0.0 | 0.3 | 0.3 | 0.0 | 0.0 | 0.1 | -0.3 | -0.8 | -0.9 |
| **MET** | 0.3 | -0.2 | 0.2 | -0.5 | -0.5 | 0.1 | -0.6 | -0.3 | -0.3 | 0.1 | -0.2 | -0.8 | 0.0 | 0.2 | -0.1 | -0.3 | -0.2 | -0.9 | -0.7 | -1.2 |
| **ASP** | 0.7 | 0.5 | 0.1 | 0.3 | 0.5 | -0.0 | 0.4 | 0.6 | 0.1 | 0.3 | -0.3 | 0.4 | -0.3 | 0.4 | 0.1 | -0.5 | -0.5 | 0.1 | -0.5 | -0.4 |
| **ASN** | 0.3 | 0.3 | -0.1 | 0.2 | 0.3 | 0.1 | 0.3 | 0.2 | -0.2 | -0.3 | -0.1 | 0.2 | 0.2 | 0.0 | 0.2 | -0.1 | -0.2 | -0.1 | -0.4 | -0.6 |
| **LEU** | 0.5 | -0.2 | 0.2 | -0.6 | -0.7 | 0.0 | -0.8 | -0.0 | -0.8 | 0.4 | 0.2 | -0.8 | 0.2 | 0.4 | 0.0 | -0.1 | -0.1 | -1.1 | -0.8 | -1.2 |
| **LYS** | 0.9 | 0.5 | 0.4 | 0.3 | 0.2 | 0.2 | 0.3 | 0.3 | 0.0 | -0.3 | 0.2 | 0.2 | 0.8 | -0.4 | 0.0 | 0.2 | 0.2 | -0.1 | -0.5 | -0.6 |
| **GLU** | 0.7 | 0.4 | 0.1 | 0.6 | 0.4 | 0.1 | 0.4 | 0.3 | 0.2 | 0.4 | 0.0 | 0.4 | -0.4 | 0.6 | 0.0 | -0.7 | -0.3 | 0.0 | -0.4 | -0.6 |
| **GLN** | 0.7 | 0.4 | 0.1 | -0.0 | 0.0 | 0.1 | 0.1 | 0.0 | -0.1 | 0.1 | 0.2 | 0.0 | 0.0 | 0.0 | 0.1 | -0.3 | -0.3 | -0.4 | -0.6 | -0.7 |
| **ARG** | 0.3 | 0.1 | 0.0 | -0.1 | -0.0 | -0.2 | 0.0 | 0.0 | -0.3 | -0.5 | -0.1 | -0.1 | 0.2 | -0.7 | -0.3 | 0.1 | -0.2 | -0.4 | -0.8 | -1.1 |
| **HIS** | 0.1 | 0.1 | -0.2 | -0.5 | -0.1 | -0.0 | -0.2 | 0.1 | -0.2 | -0.5 | -0.2 | -0.1 | 0.2 | -0.3 | -0.3 | -0.2 | -0.7 | -0.5 | -0.8 | -0.8 |
| **PHE** | -0.0 | -0.5 | -0.1 | -0.8 | -0.9 | -0.2 | -1.0 | -0.3 | -0.9 | 0.1 | -0.1 | -1.1 | -0.1 | 0.0 | -0.4 | -0.4 | -0.5 | -1.2 | -1.1 | -1.5 |
| **TYR** | -0.3 | -0.4 | -0.3 | -0.8 | -0.7 | -0.3 | -0.8 | -0.8 | -0.7 | -0.5 | -0.4 | -0.8 | -0.5 | -0.4 | -0.6 | -0.8 | -0.8 | -1.1 | -0.6 | -1.5 |
| **TRP** | -0.5 | -0.8 | -0.4 | -1.0 | -1.1 | -0.4 | -1.1 | -0.9 | -1.2 | -0.4 | -0.6 | -1.2 | -0.6 | -0.6 | -0.7 | -1.1 | -0.8 | -1.5 | -1.5 | -1.2 |

**Table 7**

| MCE/MEC | GLY | ALA | SER | CYS | VAL | THR | ILE | PRO | MET | ASP | ASN | LEU | LYS | GLU | GLN | ARG | HIS | PHE | TYR | TRP |
|---|---|---|---|---|---|---|---|---|---|---|---|---|---|---|---|---|---|---|---|---|
| **GLY** | 2.0 | 1.3 | 1.0 | 0.9 | 1.3 | 0.9 | 1.1 | 1.3 | 0.9 | 0.6 | 0.6 | 0.9 | 1.2 | 1.4 | 0.9 | 0.6 | 1.0 | 0.5 | 0.5 | 0.7 |
| **ALA** | 1.5 | 0.9 | 0.5 | 0.5 | 0.4 | 0.4 | 0.2 | 0.8 | 0.1 | 0.5 | 0.6 | 0.1 | 1.0 | 1.1 | 1.0 | 0.7 | 0.7 | -0.0 | 0.2 | -0.1 |
| **SER** | 1.0 | 1.0 | 0.4 | 0.5 | 0.7 | 0.5 | 0.7 | 0.5 | 0.4 | -0.1 | 0.1 | 0.6 | 0.5 | 0.5 | 0.7 | 0.4 | 0.1 | 0.2 | 0.1 | 0.2 |
| **CYS** | 0.6 | 0.2 | 0.2 | -1.7 | -0.2 | 0.1 | -0.6 | -0.0 | -0.5 | 0.2 | 0.3 | -0.4 | 0.4 | 0.7 | 0.1 | -0.1 | -0.5 | -0.6 | -0.3 | -0.6 |
| **VAL** | 0.7 | 0.2 | 0.3 | -0.1 | -0.4 | 0.3 | -0.5 | 0.0 | -0.5 | 0.4 | 0.3 | -0.7 | 0.7 | 0.5 | 0.3 | 0.2 | 0.1 | -0.8 | -0.4 | -0.6 |
| **THR** | 0.8 | 0.5 | 0.2 | 0.3 | 0.2 | 0.3 | 0.2 | 0.1 | 0.0 | -0.3 | -0.1 | -0.0 | 0.4 | 0.5 | 0.3 | 0.3 | 0.1 | -0.1 | 0.0 | -0.3 |
| **ILE** | 0.7 | -0.1 | 0.5 | -0.2 | -0.6 | -0.0 | -1.0 | -0.1 | -0.5 | 0.5 | 0.2 | -0.8 | 0.7 | 0.5 | 0.3 | 0.1 | 0.2 | -0.9 | -0.6 | -0.8 |
| **PRO** | 1.0 | 0.7 | 0.7 | 0.3 | 0.4 | 0.8 | 0.3 | 0.5 | 0.3 | 0.6 | 0.2 | 0.4 | 0.7 | 0.7 | 0.5 | 0.1 | 0.1 | 0.1 | -0.4 | -0.6 |
| **MET** | 0.6 | -0.0 | 0.5 | -0.1 | -0.5 | 0.1 | -0.8 | 0.1 | -0.4 | 0.3 | -0.2 | -0.7 | 0.8 | 1.0 | 0.2 | 0.1 | 0.0 | -0.9 | -0.8 | -0.8 |
| **ASP** | 0.7 | 0.7 | -0.2 | 0.5 | 0.7 | -0.1 | 0.6 | 0.3 | 0.2 | 0.0 | -0.1 | 0.2 | -0.4 | 0.6 | 0.2 | -0.7 | -0.2 | 0.2 | -0.3 | -0.1 |
| **ASN** | 0.7 | 0.6 | 0.3 | 0.8 | 0.6 | 0.3 | 0.6 | 0.2 | 0.3 | -0.3 | -0.2 | 0.5 | 0.3 | 0.3 | 0.1 | -0.2 | 0.2 | 0.2 | -0.1 | -0.5 |
| **LEU** | 0.9 | 0.0 | 0.5 | -0.2 | -0.6 | 0.2 | -0.9 | 0.0 | -0.7 | 0.7 | 0.5 | -0.9 | 0.7 | 0.8 | 0.4 | 0.3 | 0.1 | -1.0 | -0.6 | -0.8 |
| **LYS** | 1.2 | 1.0 | 0.7 | 0.8 | 0.9 | 0.6 | 0.7 | 0.9 | 0.5 | -0.0 | 0.2 | 0.5 | 1.1 | 0.3 | 0.7 | 0.7 | 0.6 | 0.3 | -0.1 | 0.1 |
| **GLU** | 1.0 | 1.0 | -0.1 | 0.7 | 0.8 | -0.2 | 0.6 | 0.4 | 0.1 | 0.4 | 0.0 | 0.4 | -0.0 | 1.0 | 0.4 | -0.3 | 0.1 | 0.2 | 0.1 | -0.1 |
| **GLN** | 0.8 | 0.5 | 0.1 | 0.3 | 0.6 | -0.2 | 0.4 | 0.3 | -0.0 | 0.0 | -0.1 | 0.1 | 0.3 | 0.4 | 0.4 | -0.0 | 0.3 | 0.1 | -0.3 | -0.3 |
| **ARG** | 0.9 | 0.6 | 0.3 | 0.4 | 0.6 | 0.1 | 0.2 | 0.1 | 0.0 | -0.3 | 0.1 | 0.2 | 0.7 | -0.1 | 0.0 | 0.1 | 0.1 | -0.1 | -0.1 | -0.3 |
| **HIS** | 0.6 | 0.5 | -0.0 | -0.7 | 0.2 | -0.1 | -0.0 | 0.1 | -0.2 | -0.5 | 0.1 | -0.1 | 0.4 | 0.2 | 0.1 | -0.1 | -0.8 | -0.4 | -0.4 | -0.7 |
| **PHE** | 0.4 | -0.1 | -0.0 | -0.7 | -0.7 | -0.1 | -1.0 | -0.2 | -1.1 | 0.1 | -0.1 | -1.0 | 0.4 | 0.2 | -0.1 | -0.2 | -0.2 | -1.2 | -0.8 | -1.2 |
| **TYR** | 0.2 | 0.2 | -0.1 | -0.2 | -0.4 | -0.2 | -0.7 | -0.6 | -0.6 | -0.4 | -0.1 | -0.7 | -0.1 | -0.0 | 0.1 | -0.5 | -0.6 | -0.9 | -0.8 | -1.0 |
| **TRP** | 0.0 | -0.6 | -0.3 | -0.6 | -0.8 | -0.3 | -0.9 | -0.9 | -1.2 | -0.0 | -0.3 | -1.2 | -0.0 | -0.0 | -0.5 | -0.7 | -0.6 | -1.3 | -0.9 | -1.1 |

**Table 8**

| ACC | GLY | ALA | SER | CYS | VAL | THR | ILE | PRO | MET | ASP | ASN | LEU | LYS | GLU | GLN | ARG | HIS | PHE | TYR | TRP |
|---|---|---|---|---|---|---|---|---|---|---|---|---|---|---|---|---|---|---|---|---|
| **GLY** | 1.5 | 0.8 | 1.1 | 0.1 | 0.6 | 0.8 | 0.5 | 0.9 | 0.4 | 1.3 | 1.0 | 0.6 | 1.3 | 1.5 | 1.1 | 1.0 | 0.9 | 0.2 | 0.2 | 0.1 |
| **ALA** | 0.8 | 0.0 | 0.6 | -0.4 | -0.6 | 0.2 | -0.7 | 0.3 | -0.4 | 1.4 | 0.9 | -0.5 | 0.9 | 1.1 | 0.8 | 0.6 | 0.3 | -0.6 | -0.6 | -0.4 |
| **SER** | 1.1 | 0.6 | 0.8 | 0.1 | 0.1 | 0.8 | -0.0 | 0.4 | -0.0 | 1.1 | 0.8 | 0.1 | 1.0 | 1.4 | 1.0 | 0.7 | 0.7 | 0.1 | 0.1 | -0.3 |
| **CYS** | 0.1 | -0.4 | 0.1 | -1.9 | -0.8 | -0.2 | -0.5 | -0.5 | -0.9 | 0.9 | 0.5 | -0.7 | 0.6 | 0.4 | 0.2 | 0.1 | -0.4 | -1.0 | -0.8 | -0.8 |
| **VAL** | 0.6 | -0.6 | 0.1 | -0.8 | -0.8 | -0.2 | -1.3 | -0.2 | -0.8 | 0.7 | 0.4 | -1.2 | 0.3 | 0.5 | 0.2 | -0.2 | -0.1 | -1.2 | -1.0 | -1.1 |
| **THR** | 0.8 | 0.2 | 0.8 | -0.2 | -0.2 | 0.5 | -0.4 | 0.2 | -0.2 | 1.2 | 0.6 | -0.3 | 1.3 | 0.9 | 0.7 | 0.3 | 0.1 | -0.3 | -0.3 | -0.5 |
| **ILE** | 0.5 | -0.7 | -0.0 | -0.5 | -1.3 | -0.4 | -1.3 | -0.2 | -1.0 | 0.7 | 0.5 | -1.3 | 0.2 | 0.3 | 0.0 | -0.2 | -0.4 | -1.4 | -1.2 | -1.4 |
| **PRO** | 0.9 | 0.3 | 0.4 | -0.5 | -0.2 | 0.2 | -0.2 | 0.6 | -0.1 | 0.7 | 0.5 | -0.2 | 0.7 | 0.7 | 0.5 | 0.2 | 0.1 | -0.5 | -0.5 | -0.8 |
| **MET** | 0.4 | -0.4 | -0.0 | -0.9 | -0.8 | -0.2 | -1.0 | -0.1 | -1.1 | 0.8 | 0.3 | -1.1 | 0.6 | 0.7 | 0.2 | -0.1 | -0.3 | -1.4 | -1.2 | -1.3 |
| **ASP** | 1.3 | 1.4 | 1.1 | 0.9 | 0.7 | 1.2 | 0.7 | 0.7 | 0.8 | 2.1 | 1.3 | 0.9 | 0.8 | 2.0 | 1.4 | 0.3 | 0.3 | 0.4 | -0.0 | 0.0 |
| **ASN** | 1.0 | 0.9 | 0.8 | 0.5 | 0.4 | 0.6 | 0.5 | 0.5 | 0.3 | 1.3 | 1.0 | 0.5 | 1.0 | 1.3 | 1.0 | 0.6 | 0.8 | 0.1 | -0.2 | -0.2 |
| **LEU** | 0.6 | -0.5 | 0.1 | -0.7 | -1.2 | -0.3 | -1.3 | -0.2 | -1.1 | 0.9 | 0.5 | -1.3 | 0.1 | 0.5 | -0.0 | -0.2 | -0.3 | -1.4 | -1.2 | -1.2 |
| **LYS** | 1.3 | 0.9 | 1.0 | 0.6 | 0.3 | 1.3 | 0.2 | 0.7 | 0.6 | 0.8 | 1.0 | 0.1 | 2.2 | 1.0 | 1.2 | 1.4 | 1.0 | 0.1 | 0.0 | -0.2 |
| **GLU** | 1.5 | 1.1 | 1.4 | 0.4 | 0.5 | 0.9 | 0.3 | 0.7 | 0.7 | 2.0 | 1.3 | 0.5 | 1.0 | 1.8 | 1.3 | 0.5 | 0.2 | 0.3 | 0.1 | -0.1 |
| **GLN** | 1.1 | 0.8 | 1.0 | 0.2 | 0.2 | 0.7 | 0.0 | 0.5 | 0.2 | 1.4 | 1.0 | -0.0 | 1.2 | 1.3 | 0.9 | 0.7 | 0.9 | -0.2 | -0.0 | -0.6 |
| **ARG** | 1.0 | 0.6 | 0.7 | 0.1 | -0.2 | 0.3 | -0.2 | 0.2 | -0.1 | 0.3 | 0.6 | -0.2 | 1.4 | 0.5 | 0.7 | 1.2 | 0.4 | -0.4 | -0.3 | -0.5 |
| **HIS** | 0.9 | 0.3 | 0.7 | -0.4 | -0.1 | 0.1 | -0.4 | 0.1 | -0.3 | 0.3 | 0.8 | -0.3 | 1.0 | 0.2 | 0.9 | 0.4 | -0.3 | -0.6 | -0.6 | -0.4 |
| **PHE** | 0.2 | -0.6 | 0.1 | -1.0 | -1.2 | -0.3 | -1.4 | -0.5 | -1.4 | 0.4 | 0.1 | -1.4 | 0.1 | 0.3 | -0.2 | -0.4 | -0.6 | -1.4 | -1.4 | -1.5 |
| **TYR** | 0.2 | -0.6 | 0.1 | -0.8 | -1.0 | -0.3 | -1.2 | -0.5 | -1.2 | -0.0 | -0.2 | -1.2 | 0.0 | 0.1 | -0.0 | -0.3 | -0.6 | -1.4 | -0.9 | -1.1 |
| **TRP** | 0.1 | -0.4 | -0.3 | -0.8 | -1.1 | -0.5 | -1.4 | -0.8 | -1.3 | 0.0 | -0.2 | -1.2 | -0.2 | -0.1 | -0.6 | -0.5 | -0.4 | -1.5 | -1.1 | -0.8 |

**Table 9**

| AEE | GLY | ALA | SER | CYS | VAL | THR | ILE | PRO | MET | ASP | ASN | LEU | LYS | GLU | GLN | ARG | HIS | PHE | TYR | TRP |
|---|---|---|---|---|---|---|---|---|---|---|---|---|---|---|---|---|---|---|---|---|
| **GLY** | 1.0 | 0.9 | 0.6 | 0.3 | 0.6 | 0.5 | 0.4 | 0.8 | 0.5 | 0.6 | 0.5 | 0.3 | 1.4 | 1.0 | 0.8 | 0.7 | 0.6 | -0.2 | 0.1 | 0.2 |
| **ALA** | 0.9 | 0.9 | 0.9 | 0.3 | -0.0 | 0.8 | -0.3 | 0.4 | -0.2 | 1.0 | 0.6 | -0.2 | 0.9 | 1.2 | 0.4 | 0.7 | 0.5 | -0.5 | -0.3 | -0.4 |
| **SER** | 0.6 | 0.9 | 1.0 | 0.3 | 0.5 | 0.7 | 0.4 | 0.5 | 0.6 | 0.8 | 0.5 | 0.3 | 1.0 | 0.9 | 0.6 | 0.4 | 0.4 | 0.1 | -0.0 | -0.4 |
| **CYS** | 0.3 | 0.3 | 0.3 | -2.4 | -0.1 | 0.1 | -0.5 | 0.3 | -0.5 | 0.4 | 0.1 | -0.7 | 0.5 | 0.6 | 0.0 | 0.1 | -0.4 | -0.8 | -0.6 | -1.4 |
| **VAL** | 0.6 | -0.0 | 0.5 | -0.1 | -0.5 | 0.4 | -0.7 | 0.2 | -0.6 | 1.1 | 0.8 | -0.8 | 0.6 | 0.7 | 0.4 | 0.3 | 0.0 | -1.0 | -0.5 | -1.3 |
| **THR** | 0.5 | 0.8 | 0.7 | 0.1 | 0.4 | 0.7 | 0.2 | 0.7 | 0.2 | 0.9 | 0.3 | 0.2 | 1.1 | 1.0 | 0.6 | 0.4 | 0.2 | -0.1 | -0.1 | -0.5 |
| **ILE** | 0.4 | -0.3 | 0.4 | -0.5 | -0.7 | 0.2 | -0.9 | 0.1 | -0.8 | 0.6 | 0.4 | -1.1 | 0.4 | 0.7 | 0.2 | 0.2 | 0.1 | -1.3 | -0.9 | -1.3 |
| **PRO** | 0.8 | 0.4 | 0.5 | 0.3 | 0.2 | 0.7 | 0.1 | 0.3 | 0.1 | 0.5 | 0.3 | 0.1 | 1.2 | 0.6 | 0.4 | 0.4 | 0.2 | -0.2 | -0.4 | -0.9 |
| **MET** | 0.5 | -0.2 | 0.6 | -0.5 | -0.6 | 0.2 | -0.8 | 0.1 | 0.1 | 0.4 | 0.1 | -0.8 | 0.5 | 0.3 | 0.4 | 0.3 | -0.3 | -1.0 | -0.8 | -0.8 |
| **ASP** | 0.6 | 1.0 | 0.8 | 0.4 | 1.1 | 0.9 | 0.6 | 0.5 | 0.4 | 1.2 | 0.4 | 0.7 | 0.5 | 1.2 | 0.5 | 0.3 | -0.0 | 0.7 | 0.2 | 0.2 |
| **ASN** | 0.5 | 0.6 | 0.5 | 0.1 | 0.8 | 0.3 | 0.4 | 0.3 | 0.1 | 0.4 | 0.6 | 0.4 | 1.2 | 0.9 | 0.3 | 0.4 | -0.3 | 0.2 | 0.1 | -0.5 |
| **LEU** | 0.3 | -0.2 | 0.3 | -0.7 | -0.8 | 0.2 | -1.1 | 0.1 | -0.8 | 0.7 | 0.4 | -0.9 | 0.2 | 0.7 | 0.3 | 0.0 | -0.2 | -1.2 | -1.1 | -1.6 |
| **LYS** | 1.4 | 0.9 | 1.0 | 0.5 | 0.6 | 1.1 | 0.4 | 1.2 | 0.5 | 0.5 | 1.2 | 0.2 | 1.7 | 0.8 | 0.6 | 1.0 | 0.6 | 0.4 | 0.1 | 0.4 |
| **GLU** | 1.0 | 1.2 | 0.9 | 0.6 | 0.7 | 1.0 | 0.7 | 0.6 | 0.3 | 1.2 | 0.9 | 0.7 | 0.8 | 1.7 | 0.8 | 0.3 | 0.5 | 0.4 | 0.1 | -0.2 |
| **GLN** | 0.8 | 0.4 | 0.6 | 0.0 | 0.4 | 0.6 | 0.2 | 0.4 | 0.4 | 0.5 | 0.3 | 0.3 | 0.6 | 0.8 | 1.2 | 0.6 | 0.6 | -0.2 | -0.1 | -0.4 |
| **ARG** | 0.7 | 0.7 | 0.4 | 0.1 | 0.3 | 0.4 | 0.2 | 0.4 | 0.3 | 0.3 | 0.4 | 0.0 | 1.0 | 0.3 | 0.6 | 1.0 | 0.4 | -0.2 | -0.3 | -0.4 |
| **HIS** | 0.6 | 0.5 | 0.4 | -0.4 | 0.0 | 0.2 | 0.1 | 0.2 | -0.3 | -0.0 | -0.3 | -0.2 | 0.6 | 0.5 | 0.6 | 0.4 | -0.2 | -0.5 | -0.4 | -0.9 |
| **PHE** | -0.2 | -0.5 | 0.1 | -0.8 | -1.0 | -0.1 | -1.3 | -0.2 | -1.0 | 0.7 | 0.2 | -1.2 | 0.4 | 0.4 | -0.2 | -0.2 | -0.5 | -1.2 | -1.0 | -1.5 |
| **TYR** | 0.1 | -0.3 | -0.0 | -0.6 | -0.5 | -0.1 | -0.9 | -0.4 | -0.8 | 0.2 | 0.1 | -1.1 | 0.1 | 0.1 | -0.1 | -0.3 | -0.4 | -1.0 | -0.7 | -1.1 |
| **TRP** | 0.2 | -0.4 | -0.4 | -1.4 | -1.3 | -0.5 | -1.3 | -0.9 | -0.8 | 0.2 | -0.5 | -1.6 | 0.4 | -0.2 | -0.4 | -0.4 | -0.9 | -1.5 | -1.1 | -0.8 |

**Table 10**

| ACE/AEC | GLY | ALA | SER | CYS | VAL | THR | ILE | PRO | MET | ASP | ASN | LEU | LYS | GLU | GLN | ARG | HIS | PHE | TYR | TRP |
|---|---|---|---|---|---|---|---|---|---|---|---|---|---|---|---|---|---|---|---|---|
| **GLY** | 1.2 | 1.2 | 1.0 | 0.7 | 0.7 | 0.6 | 0.7 | 1.1 | 0.5 | 0.1 | 0.1 | 0.7 | 1.7 | 1.5 | 1.1 | 0.8 | 0.9 | 0.5 | 0.3 | 0.6 |
| **ALA** | 1.0 | 0.0 | 0.9 | -0.2 | -0.6 | 0.6 | -0.8 | 0.6 | -0.5 | 1.0 | 0.9 | -0.5 | 1.2 | 1.3 | 0.8 | 0.6 | 0.3 | -0.7 | -0.6 | -0.6 |
| **SER** | 1.1 | 0.9 | 0.7 | 0.3 | 0.2 | 0.8 | 0.3 | 0.7 | 0.3 | 0.7 | 0.6 | 0.1 | 0.9 | 1.3 | 0.6 | 0.4 | 0.4 | -0.0 | -0.2 | 0.0 |
| **CYS** | 0.3 | -0.3 | 0.4 | -2.6 | -0.8 | -0.2 | -0.9 | -0.4 | -0.9 | 0.5 | 0.2 | -0.9 | 0.1 | 1.1 | 0.4 | -0.1 | -0.5 | -1.3 | -1.1 | -1.4 |
| **VAL** | 0.4 | -0.3 | 0.4 | -0.5 | -1.1 | 0.2 | -1.3 | 0.0 | -0.7 | 0.6 | 0.5 | -1.1 | 0.8 | 0.5 | 0.3 | -0.0 | -0.2 | -1.3 | -1.0 | -1.2 |
| **THR** | 0.8 | 0.3 | 0.5 | -0.3 | -0.1 | 0.7 | -0.4 | 0.3 | -0.3 | 0.8 | 0.3 | -0.3 | 0.8 | 0.8 | 0.6 | 0.4 | -0.2 | -0.5 | -0.5 | -0.6 |
| **ILE** | 0.4 | -0.5 | 0.7 | -0.4 | -1.1 | 0.0 | -1.5 | -0.2 | -1.0 | 0.8 | 0.4 | -1.3 | 0.5 | 0.5 | 0.1 | -0.1 | -0.2 | -1.4 | -1.2 | -1.3 |
| **PRO** | 1.0 | 0.7 | 0.7 | 0.3 | 0.3 | 0.7 | 0.2 | 0.4 | 0.3 | 1.1 | 0.6 | 0.2 | 1.2 | 0.8 | 0.5 | 0.5 | -0.2 | -0.0 | -0.4 | -0.8 |
| **MET** | 0.2 | -0.7 | 0.4 | -0.5 | -0.8 | -0.2 | -1.2 | -0.1 | -0.9 | 0.5 | -0.0 | -1.2 | 0.9 | 1.0 | 0.8 | -0.2 | 0.0 | -1.4 | -1.0 | -1.3 |
| **ASP** | 1.3 | 0.8 | 0.8 | 0.9 | 1.0 | 1.2 | 0.4 | 1.0 | 0.9 | 1.6 | 0.7 | 0.8 | 0.5 | 1.9 | 1.3 | 0.3 | 0.3 | 0.5 | -0.2 | 0.0 |
| **ASN** | 0.9 | 0.7 | 0.7 | 0.3 | 0.6 | 0.4 | 0.3 | 0.4 | 0.4 | 0.7 | 0.5 | 0.3 | 1.1 | 1.1 | 0.8 | 0.4 | 0.4 | -0.0 | 0.1 | -0.2 |
| **LEU** | 0.5 | -0.5 | 0.3 | -0.6 | -1.2 | 0.1 | -1.4 | -0.2 | -0.8 | 1.1 | 0.5 | -1.3 | 0.5 | 0.7 | 0.3 | -0.0 | -0.4 | -1.4 | -1.2 | -1.2 |
| **LYS** | 1.2 | 1.0 | 1.0 | 0.7 | 0.3 | 0.9 | -0.0 | 0.8 | 0.1 | 0.7 | 1.1 | 0.1 | 1.5 | 0.9 | 0.8 | 1.7 | 0.9 | -0.2 | -0.1 | -0.3 |
| **GLU** | 1.3 | 1.0 | 0.9 | 0.6 | 0.4 | 0.8 | 0.2 | 0.9 | 0.5 | 1.8 | 0.9 | 0.3 | 0.8 | 1.6 | 1.6 | 0.3 | 0.5 | -0.0 | -0.2 | 0.1 |
| **GLN** | 0.7 | 0.7 | 0.7 | 0.1 | 0.1 | 0.6 | -0.2 | 0.3 | 0.3 | 0.9 | 0.5 | -0.1 | 1.3 | 1.1 | 1.2 | 0.7 | 0.3 | -0.3 | -0.4 | -0.7 |
| **ARG** | 1.0 | 0.6 | 0.7 | -0.0 | 0.1 | 0.2 | -0.2 | 0.2 | 0.1 | 0.0 | 0.6 | -0.3 | 1.2 | 0.1 | 0.4 | 0.7 | 0.4 | -0.5 | -0.5 | -0.6 |
| **HIS** | 0.9 | -0.0 | 0.3 | -0.9 | -0.2 | 0.3 | -0.2 | 0.2 | -0.1 | 0.2 | -0.1 | -0.1 | 1.0 | 0.3 | 1.2 | 0.3 | -0.4 | -0.7 | -0.7 | -0.1 |
| **PHE** | 0.3 | -0.5 | 0.1 | -0.7 | -1.2 | -0.2 | -1.4 | -0.3 | -1.1 | 0.4 | 0.5 | -1.3 | 0.1 | 0.4 | -0.2 | -0.3 | -0.6 | -1.7 | -1.2 | -1.4 |
| **TYR** | 0.4 | -0.1 | 0.5 | -0.3 | -0.7 | -0.1 | -0.9 | -0.5 | -1.0 | -0.0 | -0.0 | -0.9 | 0.2 | 0.1 | -0.5 | -0.5 | -0.4 | -1.2 | -0.8 | -1.1 |
| **TRP** | -0.1 | -0.5 | 0.0 | -0.9 | -1.0 | -0.3 | -1.3 | -1.0 | -1.3 | 0.2 | -0.3 | -1.2 | 0.1 | 0.1 | 0.0 | -0.7 | -0.9 | -1.5 | -1.2 | -1.3 |